# Strong magnetic coupling in the hexagonal $R_5Pb_3$ compounds ($R$ = Gd-Tm)

Andrea Marcinkova[1], Clarina de la Cruz[2], Joshua Yip[1], Liang L. Zhao[1], Jiakui K. Wang[1], E. Svanidze[1] and E. Morosan[1]

[1]*Department of Physics and Astronomy, Rice University, Houston, Texas 77005, USA*
[2]*Oak Ridge National Laboratory, Oak Ridge, Tennessee 37831, USA*

ABSTRACT

We have synthesized $R_5Pb_3$ ($R$ = Gd-Tm) compounds in polycrystalline form and performed structural analysis, magnetization, and neutron scattering measurements. For all $R_5Pb_3$ reported here the Weiss temperatures $\theta_W$ are several times smaller than the ordering temperatures $T_{ORD}$, while the latter are remarkably high ($T_{ORD}$ up to 275 K for $R$ = Gd) compared to other known $R$-$M$ binaries ($M$ = Si, Ge, Sn and Sb). The magnetic order changes from ferromagnetic in $R$ = Gd, Tb to antiferromagnetic in $R$ = Dy-Tm. Below $T_{ORD}$, the magnetization measurements together with neutron powder diffraction show complex magnetic behavior and reveal the existence of up to three additional phase transitions. We believe this to be a result of crystal electric field effects responsible for high magnetocrystalline anisotropy. The $R_5Pb_3$ magnetic unit cells for $R$ = Tb-Tm can be described with incommensurate magnetic wave vectors with spin modulation either along the *c* axis in $R$ = Tb, Er and Tm or within the *ab*-plane in $R$ = Dy and Ho.

## I. INTRODUCTION

The discovery of intermetallic compounds with the chemical formula $R_5M_3$ ($R$ = rare earth, $M$ = Si, Ge, Sn, Sb and Bi) attracted attention in condensed matter physics because of their rich structural and physical properties. $R_5Si_3$ compounds ($R$ = La–Nd) crystallize in the $Cr_5B_3$-type tetragonal structure[1, 2] with space group (SG) $I_4/mcm$, while the $R$ = Gd–Lu, Y, members of this series crystallize in the $Mn_5Si_3$-type hexagonal structure with SG $P6_3/mcm$.[3, 4] All $R_5Bi_3$ compounds crystallize in the orthorhombic $Y_5Bi_3$-type structure with SG *Pnma*.[2, 5, 6] $R_5M_3$ (where $R$ = La-Nd, Gd-Lu and $M$ = Ge, Sn, Sb, Pb) adopt a hexagonal $P6_3/mcm$ structure, in which the $R$ atoms ($R_1$, $R_2$) occupy two inequivalent crystallographic sites, with 2(3) $R_1(R_2)$/f.u. in the $4d(6g)$ positions. Thus one can expect different magnetic coupling between the two different sublattices and, hence, complex magnetic ordering.

Limited physical properties characterization is reported for the hexagonal $R_5M_3$ ($M$ = Si, Ge, Sn and Sb) family of compounds. $Yb_5Si_3$ is reported to form with mixed-valence Yb ions, showing antiferromagnetic order below the Néel temperature $T_N = 1.6$ K. $Ho_5Si_3$ undergoes two magnetic transitions at low temperatures: first, antiferromagnetic ordering at $T_N = 24$ K, and second, a spin reorientation transition at $T_2 \sim 8$ K. Field-dependent magnetization data reveals a metamagnetic transition at $H_c = 2.2$ T. Later studies by Canepa *et al.*[7] on $R_5Si_3$ samples with $R$ = Y, Ce, Pr, Nd, Sm, Gd, Tb, Yb, and Lu presented temperature-dependent ($T$ = 4 - 300 K) resistivity data. However, detailed information on the magnetic properties of the $R_5Si_3$ is still lacking.

The $R_5Ge_3$ compounds with $R$ = Pr, Gd, Tb, Ho, and Er show antiferromagnetic order with Néel temperatures between $T_N = 10$ and 85 K and ferrimagnetic (FI) order in $Nd_5Ge_3$ and $Ce_5Ge_3$.[8, 9] As expected, $La_5Ge_3$ and $Y_5Ge_3$ are Pauli paramagnets. No other magnetic transitions were observed below the ordering temperature in the magnetic $R_5Ge_3$ series. In 2004, Tsutaoka *et al.*[10] presented the magnetic and transport properties of $Gd_5Ge_3$ and $Tb_5Ge_3$ single crystals. $Gd_5Ge_3$ was found to be antiferromagnetic below $T_N = 76$ K, with a spin reorientation transition at $T_{SR} \sim 52$ K. $Tb_5Ge_3$ shows an antiferromagnetic order below $T_N = 79$ K, with large magnetic anisotropy.[11] The antimonides $R_5Sb_3$ with $R$ = Pr, Nd, Gd-Tm show antiferromagnetic order, with Néel temperatures up to 109 K for $R$ = Gd.[2, 12] Not surprisingly, $La_5Sb_3$ and $Y_5Sb_3$ display a temperature-independent susceptibility.[2]

Magnetization measurements carried out on $R_5M_3$ ($R$ = rare earth, $M$ = Si, Ge, Sn, and Sb) compounds revealed the presence of strong crystal field effects (CEF). As a result, these compounds present complex magnetic behavior, mainly due to exchange interaction between two inequivalent $R$ atomic sites. Such a behavior was confirmed by neutron powder diffraction (NPD) in some of the $R_5M_3$ systems. The magnetic configuration was found to be conical-spiral for $Tb_5Sb_3$,[13] flat-spiral for $Tb_5Ge_3$[14], and amplitude sine-modulated for $Ho_5Si_3$,[1] $Ho_5Sb_3$,[15] $Dy_5Sb_3$,[16] $Tb_5Sn_3$,[17] $Er_5Sn_3$[18] and $Ho_5Sn_3$.[19] To better understand the complex interplay between the exchange interactions and CEF in the $R_5M_3$ compounds, a detailed study on their anisotropic thermodynamic and transport properties is needed.

In this manuscript, we performed such a study of the structural and magnetic properties on the $R_5Pb_3$ compounds ($R$ = Gd-Tm), from a combination of magnetization and NPD measurements. We investigated the influence of different rare earth metals on the crystal structure and the magnetism.

Lattice parameters and the unit cell volume decrease monotonically with decreasing rare earth ionic radii, according to the expected lanthanide contraction.[20] NPD reveals the anisotropy in the $R_5Pb_3$ systems ($R$ = Tb-Tm) and multiple magnetic transitions in the ordered state, with incommensurate magnetic wave vectors $\mathbf{k_m}$ associated with the ordered state down to the lowest measured temperature ($T$ = 4 K). The multiple magnetic transitions observed in NPD are further confirmed by magnetization measurements. Remarkably, after $Gd_5Si_4$ ($T_{ORD}$ = 340 K),[21] $R_5Pb_3$ systems have the highest observed ordering temperatures $T_{ORD}$ compared to other known $R$-$M$ ($M$ = Si, Ge, Sn, and Sb) binary compounds. For example, $Gd_5Pb_3$ orders at 275 K, which is much higher than $Gd_5Ge_4$ ($T_{ORD}$ = 127 K),[n2] $Gd_5Sn_4$ ($T_{ORD}$ = 80 K),[22] $Gd_5Sn_3$ ($T_{ORD}$ = 68 K),[10] $Gd_5Si_3$ ($T_{ORD}$ = 55 K),[7] $Gd_5Ge_3$ ($T_{ORD}$ = 52 K).[10] A striking result of the high anisotropy in the $R_5Pb_3$ compounds is that the Weiss temperatures $\theta_W$ are up to five times smaller than the ordering temperatures $T_{ORD}$. This was also the case in the high $T_{ORD}$ Gd silicates, for example in GdSi, which has a ordering temperature ($T_{ORD}$ = 78 K) four times larger than the Weiss temperature ($\theta_W$ = 20 K).[23]

## II. MATERIALS AND METHODS

Polycrystalline $R_5Pb_3$ ($R$ = Gd-Tm) samples were prepared from rare earth ingots (Ames lab 99.999%) and Pb (Alfa Aesar; 99.98%) in a molar ratio of 5:3.5, with excess Pb used to compensate for mass losses during heating. Arc melting of the samples was performed on a water-cooled Copper hearth under a purified argon atmosphere. To ensure homogeneity, the samples were turned over and re-melted several times. Initial phase analysis was done using x-ray powder diffraction on a Rigaku D/max ULTIMA II diffractometer with Cu Kα radiation source. The $R_5Pb_3$ ($R$ = Gd-Tm) turned out to be extremely air-sensitive, so the x-ray powders were covered by a layer of amorphous mineral oil. Rietveld analysis[24] of the powder diffraction data was done using the GSAS/EXPGUI suite of programs.[25] A pseudo-Voigt function was used to describe the peak shape for all data. As an example, the Rietveld fit of the x-ray powder diffraction for $Ho_5Pb_3$ is shown in Fig. 1. The presence of the mineral oil as a protective layer can be observed as a broad peak at low angles ($2\theta$ = 10° - 20°), together with up to 3 wt% of Pb impurity in all polycrystalline $R_5Pb_3$ samples.

For the NPD experiments, roughly 5 g of each polycrystalline sample was held in a cylindrical vanadium container in a top-loading closed-cycle refrigerator and studied using the HB-2A powder diffractometer at the High Flux Isotope Reactor of Oak Ridge National Laboratory.[26] Data from HB-2A were collected with neutron wavelengths λ = 1.54 Å and λ = 2.41 Å, by (115) and (113) reflections from a vertically-focusing Ge monochromator. The data were acquired in the temperature range of 4 – 220 K by scanning the detector array consisting of 44 $^3$He tubes in two segments to cover the angle $2\theta$ range of 4°–150° in steps of 0.05°. Overlapping detectors for the given steps were used to average the counting efficiency of each detector. The shorter wavelength, which gives a greater intensity and higher $Q$ coverage, was used to investigate the crystal structures in this low temperature regime, while the longer wavelength gives lower $Q$ coverage and greater resolution that was important for investigating the magnetic structures of these materials. More details about the HB-2A instrument and data collection strategies can be found in Ref. [27]. The NPD data were analyzed using the Rietveld refinement program FULLPROF[27] and the representational analysis software *SARAh*.[28]

Powder diffraction patterns reveal that all samples have a hexagonal structure with SG $P6_3/mcm$, in which the rare earth atoms occupy two inequivalent crystallographic 4*d* and 6*g* sites, located at (1/3, 2/3, 0) and ($x_R$, 0, 1/4). The atoms in the 4*d* position (*R1* in Fig. 2c) are linked into linear chains along the *c* axis, while the atoms in the 6*g* position (*R2* in Fig. 2c) form face-sharing octahedral chains along *c*. This structure is similar to that of the $R_5M_3$ analogues ($M$ = Si, Ge, Sn). Both *a* (squares, left axis in Fig. 2a) and *c* (triangles, right axis in Fig. 2a) lattice parameters, as well as the unit cell volume (circles, right axis in Fig. 2b), decrease as expected as *R* changes from Gd towards Tm in the $R_5Pb_3$ series. From Rietveld refinements, the distances between the nearest (NN, $R_1$-$R_1$, $R_2$-$R_2$) and next-nearest (NNN, $R_1$-$R_2$) *R* neighbors were calculated. The shortest *R-R* distances $d_{R1-R1}$ = 3.2-3.4 Å were found between $R_1$ atoms situated in the quasi-one dimensional chains, parallel to the *c* axis.

Zero field-cooled (ZFC) and field-cooled (FC) DC magnetic susceptibilities were measured using a Quantum Design (QD) Magnetic Property Measurement System (MPMS). The temperature dependence of the susceptibility for all samples was measured in $H$ = 0.1 T, with additional fields up to $H$ = 7 T measured for a subset of the samples. The antiferromagnetic (AFM) transition temperatures $T_N$ and the lower-temperature transitions for the AFM compounds were determined from the peaks in $d(MT)/dT$.[29] In the case of ferromagnetic (FM) order, $T_C$ and all other transition temperatures were determined from susceptibility derivatives $dM/dT$. To further characterize the magnetic behavior of $R_5Pb_3$ samples, the field-dependent magnetization measurements $M(H)$ were performed at $T$ = 2 K. All $R_5Pb_3$ samples show metamagnetic phase transitions, with the critical field values determined from the maxima in $dM/dH$.

## III. RESULTS

### 3.1 $Gd_5Pb_3$

Fig. 3a shows the ZFC (closed symbols) temperature dependence of the magnetic susceptibility *M/H* for $Gd_5Pb_3$ in applied magnetic field $H = 0.1$ T (black circles), 1 T (red triangles) and 7 T (blue squares). The Curie-Weiss law is followed at temperatures above 300 K, as revealed by the linear inverse susceptibility H/M (open circles, right axis, Fig. 3a). The linear fit above ~300 K yields a Weiss temperature $\theta_W = 158.7$ K and effective magnetic moment $\mu_{eff}^{exp} \approx 7.46\,\mu_B$ / $Gd^{3+}$, close to the theoretical value $\mu_{eff}^{theory} \approx 7.94\,\mu_B$ for $Gd^{3+}$. The positive $\theta_W$ value is indicative of ferromagnetic coupling, consistent with the shape of the magnetic susceptibility, abruptly increasing below $T_{ORD} \sim 275$ *K*. More interestingly, $\theta_W$ is nearly half of the ordering temperature $T_{ORD} = 275$ K, with $T_{ORD}$ determined from the minimum in *dM/dT* for $H = 0.1$ T, as shown in Fig.3b. Below $T_{ORD}$ another magnetic transition is evident at $T_2 = 85.5$ K, as seen in the high field M(T) data (inset, Fig. 3a) and also revealed by the magnetization derivative *dM/dT* (inset, Fig. 3b).[29] The transition temperatures $T_{ORD}$ and $T_2$ do not change with increasing magnetic field up to $H = 7$ T

Compared to others $R_5Pb_3$ compounds, the *M(H)* isotherm for $Gd_5Pb_3$ (Fig. 4) shows almost linear behavior in the applied magnetic field of $0.2\text{ T} < H < 7\text{ T}$. However, within our field range the maximum magnetization (~ 1.5 $\mu_B/Gd^{3+}$) is much smaller than the $Gd^{3+}$ saturated moment $\mu_{sat} = 7\,\mu_B/Gd^{3+}$. The steep M(H) increase for fields H < 0.2 T, is reminiscent of the anisotropic behavior in some Gd intermetallic compounds,[30] suggesting that a helical (or more complex) magnetic moment configuration in $Gd_5Pb_3$ may be responsible for both the M(H) shape and the small *M* values reached with $H = 7$ T. No magnetic hysteresis is observed (Fig. 4).

No NPD has been collected on $Gd_5Pb_3$ due to its very high absorption cross-section for neutrons. For example, $Gd_2O_3$ is used as a neutron beam stopper.[31]

### 3.2 $Tb_5Pb_3$

The ZFC (full circles) and FC (full triangles) temperature dependence of the magnetic susceptibility *M/H* (left axis) and the inverse susceptibility *H*/*M* (right axis) is shown in Fig. 5a and reveals two magnetic transitions, first at $T_{ORD} = 215.3$ K and second at $T_2 = 69.4$ K. The values of $T_{ORD}$ and $T_2$ were determined from peaks in *dM/dT*. The inverse magnetic susceptibility shows linear temperature dependence above 300 K, indicative of Curie-Weiss behavior. The linear fit *H*/*M* above ~ 300 K yields a Weiss temperature $\theta_W = 96.4$ K. This is less than half of $T_{ORD}$, a likely consequence of strong crystal field effects, or anisotropic coupling, or both. The positive $\theta_W$ value indicates the dominance of the ferromagnetic coupling between the $Tb^{3+}$ magnetic ions. The effective magnetic moment is $m_{eff}^{exp} \approx 9.66\ \mu_B/Tb^{3+}$, which is consistent with the theoretical value calculated for the $Tb^{3+}$ free ion, $\mu_{eff}^{theory} \approx 9.72\ \mu_B$. The $T = 2$ K $M(H)$ isotherm (black squares, Fig. 5b) shows small hysteresis below $H \sim 2.2$ T. Considering the lack of the secondary phases in the neutron data shown below, the observed small hysteresis in $Tb_5Pb_3$ measured at $T = 2$ K can be ascribed to a small ferromagnetic component. Within our field range, the maximum magnetization ($\sim 2\ \mu_B/Tb^{3+}$) is much smaller than the $Tb^{3+}$ saturated moment $\mu_{sat} = 9\ \mu_B/Tb^{3+}$. The low measured magnetization values $M(H) \leq 2\ \mu_B$ could be a consequence of the crystal field anisotropy.

The magnetic order in $Tb_5Pb_3$ is further investigated by the NPD data collected at $T = 300$, 150, 90, 70, 60, 50 K using a wavelength $\lambda = 1.54$ Å, and at $T = 4$ K using $\lambda = 2.41$ Å (Fig. 6). In the paramagnetic state ($T = 300$ K bottom curve in Fig. 6), only the nuclear peaks are visible. As the temperature is decreased through $T = 150$ K, extra intensity on the nuclear (2,1,1) Bragg peak develops at Q = 2.29 $Å^{-1}$ for all lower temperatures. Additional magnetic peaks are observed below $T = 70$ K, in the range $Q = 1 - 2.5\ Å^{-1}$, with intensities varying as a function of temperature. Using a larger wavelength $\lambda = 2.41$ Å, a $(0,0,l)$ magnetic peak with $l = 1.52$ is revealed at $T = 4$ K (inset, Fig. 6), suggesting that the spins order along $c$ axis at this temperature.

A careful comparison of the $T = 300$ K and 150 K data (inset, Fig. 7) reveals weak magnetic contributions to the nuclear (1,0,2), (1,1,1), and (2,1,1) Bragg reflections. This is indicative of weak ferromagnetic coupling of the $R$ spins, which is consistent with the ferromagnetic ordering observed from magnetization around $T_{ORD} = 215.3$ K (Fig. 5a). All magnetic Bragg reflections were indexed with a hexagonal cell $a_M = a_N$, $c_M = c_N$, where the subscript $M$ denotes magnetic and $N$ the nuclear cell. The temperature evolution of the magnetic (2,1,1)+$\mathbf{k_m}$ Bragg reflection was measured in the temperature range $T = 150$ K - 260 K, and its intensity is shown in Fig. 7 (black squares). The intensity of this peak increases in the ordered state, resulting in an estimate of the ordering temperature $T_{ORD} = 211$ K, taken as the intercept of the two solid lines in Fig. 7. This is in good agreement with the magnetization data shown in Fig. 5a.

Upon further cooling, the new magnetic peaks developing below $T = 70$ K point to an AFM state in this temperature range. Indexing these magnetic peaks suggests that the magnetic unit cell is almost twice as large as the nuclear unit cell with $a_M = a_N$ and $c_M = 1.88c_N$. The corresponding propagation magnetic vector below $T = 60$ K is $\mathbf{k_{m1}} = (0,\ 0,\ 0.535)$, Fig. 8a. At 4 K, the incommensurability changes, with the corresponding magnetic vector $\mathbf{k_{m2}} = (0, 0, 0.520)$, Fig. 8b. A detailed analysis of the magnetic structure in $Tb_5Pb_3$ is underway, to be published elsewhere.[32] The peak at $Q = 2.29$ Å$^{-1}$ persists throughout the AFM state (Fig. 8), making it unclear whether it is intrinsic to the AFM order or if it is associated with a FM component within the AFM state.

### 3.3 $Dy_5Pb_3$

Fig. 9a shows the ZFC (circles) and FC (triangles) temperature-dependent magnetic susceptibility $M/H$ (left axis) and the inverse susceptibility $H/M$ (open symbols, right axis), measured in an applied field $H = 0.1$ T. By contrast with the $R$ = Gd and Tb compounds, where $M(T)$ diverged at $T_{ORD}$ (Figs., 3a and 5a), in $Dy_5Pb_3$ the magnetic order is indicated by a peak close to 160 K. The shape of the magnetic susceptibility and the indistinguishable ZFC/FC $M(T)$ close to this temperature suggest that the transition corresponds to AFM order, with the Néel temperature $T_N$ determined from $d(MT)/dT$[23] to be $T_N = 160.6$ K. However, the upturn in the magnetization and the additional peaks at lower temperatures reveal very complex magnetic behavior in $Dy_5Pb_3$. Three local magnetization maxima at $T_2 = 75.3$ K, $T_3 = 43.3$ K and $T_4 = 15.4$ K are determined from the $d(MT)/dT$, with small ZFC/FC irreversibility observed below $T_2$. The spin reorientation transitions at $T_2$-$T_4$ might also be associated with a small ferromagnetic component at low temperatures. The inverse magnetic susceptibility displayed in Fig. 9a (right axis) is linear above 220 K, indicative of Curie-Weiss behavior. From the linear fit of $H/M$ above $T = 220$ K, the resulting Weiss temperature is $\theta_W = 36.4$ K, with the effective moment $\mu_{eff}^{exp} \approx 10.38\ \mu_B/Dy^{3+}$, consistent with the theoretical value $\mu_{eff}^{theory} \approx 10.63\ \mu_B/Dy^{3+}$. The positive $\theta_W$ value indicates FM coupling between $Dy^{3+}$ ions. It is very likely then that strong CEF effects result in the AFM order, and are responsible for the remarkably large ordering temperature (almost five times) compared to $\theta_W$.

The $M(H)$ isotherm measured at $T = 2$ K (Fig. 9b) displays one broad metamagnetic transition at $H_c = 5.9$ (6.25) T for increasing (decreasing) applied field, as determined from the peaks position in the $dM/dH$ plot (inset, Fig. 9b). The magnetization in the maximum applied field $H = 7$ T (~5.8 $\mu_B$ / $Tb^{3+}$) is close to half of the $Dy^{3+}$ saturated moment $\mu_{sat} = 10\ \mu_B/Dy^{3+}$. It is therefore likely that more metamagnetic transitions occur in higher magnetic fields. Small hysteresis can be observed for the whole measured field range, a likely result of the FM component at low temperatures, as suggested by the ZFC/FC irreversibility in $M(T)$ data (Fig. 9a).

Neutron diffraction measurements on Dy-containing samples are inherently difficult given that Dy is a strong neutron absorber. However, measurements on $Dy_5Pb_3$ did reveal two magnetic peaks between $T$ = 30 K and 4 K, marked by the vertical arrows in Fig. 10a. These magnetic peaks were identified by comparison with the measurement in the paramagnetic state $T = 250$ K (bottom curve, Fig. 10a). The order parameter for $Dy_5Pb_3$, measured on the (-3,1,1)+$\mathbf{k_m}$ magnetic Bragg reflection in the temperature range $T = 4$ - 95 K (Fig. 10b) reveals the third magnetic transition at $T_3$ = 39 K. This is close to the value determined from magnetization $T_3$ = 43.3 K (Fig. 9a). The magnetic peaks are consistent with a magnetic wavevector $\mathbf{k_m}$ = (0, 0.301, 0), determined from the Lebail fit of the $T$ = 4 K NPD data (Fig. 11).

### 3.4 $Ho_5Pb_3$

Fig. 12a shows the temperature-dependent magnetic susceptibility $M/H$ (black circles, left axis) and the inverse susceptibility $H/M$ (open circles, right axis) for $Ho_5Pb_3$ in an applied field $H = 0.1$ T. No ZFC/FC irreversibility was registered down to 1.8 K, which, together with the weak peak around 110 K, suggest AFM order in this compound. The Néel temperature $T_N$ was determined from $d(MT)/dT$ to be $T_N = 111$ K. Upon further cooling, two local magnetization maxima at $T_2 = 23.7$ K and $T_3 = 8.03$ K are determined from the $d(MT)/dT$. The inverse magnetic susceptibility displayed in Fig. 12 (right axis) is linear above 110 K, indicative of Curie-Weiss behavior. From the linear fit of $H/M$ above $T = 110$ K, the resulting Weiss temperature is $\theta_W = 30.4$ K, with the effective moment $m_{eff}^{exp} \approx 10.95\ \mu_B/Ho^{3+}$ close to the theoretical value $\mu_{eff}^{theory} \approx 10.61\ \mu_B/Ho^3$. The positive $\theta_W$ value indicates FM coupling between $Ho^{3+}$ ions. However, the evidence for magnetic order from M(T) data points to AFM order, suggesting possible anisotropic spin coupling and strong CEF effects, as well as the remarkably large ordering temperature $T_{ORD}$ compared to $\theta_{W,}$ $T_{ORD}/\theta_W$ ~4.

The $T$ = 2 K $M(H)$ isotherm (Fig. 12b) shows two metamagnetic transitions up to $H$ = 7 T, also consistent with spin-flop transitions in the AFM state. However, within our field range the maximum magnetization (~3 $\mu_B/Ho^{3+}$) is less than half the $Ho^{3+}$ saturated moment $\mu_{sat}(Ho^{3+}) = 10.6\ \mu_B$. It is therefore likely that more metamagnetic transitions occur in higher magnetic fields, or it could be a consequence of the crystal field anisotropy.

Fig. 13 shows the temperature evolution of the (3,-1,0)+$\mathbf{k_{m1}}$, (1,0,0)+$\mathbf{k_{m2}}$, and (0,1,0)+$\mathbf{k_{m2}}$+$\mathbf{k_{m3}}$ as the most intense Ho5Pb3 magnetic Bragg peaks for the respective temperature intervals. The order parameter for $Ho_5Pb_3$ measured on the (3,-1,0)+$\mathbf{k_{m1}}$ magnetic Bragg reflection in the temperature range $T = 70$ - 120 K reveals the ordering temperature $T = 110.5$ K, taken as the intercept of the two solid lines in Fig. 13a. This is close to the value determined from magnetization $T_{ORD} = 111$ K. The (1,0,0)+$\mathbf{k_{m2}}$ and (0,1,0)+$\mathbf{k_{m2}}$+$\mathbf{k_{m3}}$ magnetic Bragg reflections measured in the temperature range $T$ = 4 - 40 K (Fig. 13b) were used to determine the magnetic transition temperatures $T_2$ ~ 28.6 K and $T_3$ ~ 8.7 K. These transition temperatures are close to the values determined from the magnetic susceptibility in Fig. 12a.

A comparison of the NPD patterns for $Ho_5Pb_3$ collected at $T$ = 150 and 50 K (Fig. 14) for $Q$ = 1.9-2.3 $Å^{-1}$ points to an AFM state in this temperature range. Indexing these magnetic peaks suggests that the magnetic unit cell is more than three times larger than the nuclear unit cell with $a_M = 3.46a_N$ and $c_M = c_N$. The corresponding propagation magnetic vector determined from the Lebail fit to the $T$ = 50 K NPD data (Fig. 14) is $\mathbf{k_{m1}}$ = (0, 0.295, 0). The ($h$,0+$\mathbf{k}$,0) magnetic peak with $h$ = 1 and $k$ = 0.295 at $T$ = 50 K (left inset, Fig. 14) suggests that the Ho spins order within the $ab$-plane at this temperature. With the second magnetic transition at $T_2$ = 28.6 K, the incommensurability of the magnetic vector changes to $\mathbf{k_{m2}}$ = (0, 0.288, 0) (orange, Fig. 15a). Below $T_3$ = 8.7 K, the additional AFM ordering along the $b$ axis is indexed using $\mathbf{k_{m3}}$ = (0, 0.5, 0) (blue, Fig 15b). At $T$ = 4 K, all magnetic Bragg reflections are indexed using the same two magnetic propagation vectors $\mathbf{k_{m2}}$ and $\mathbf{k_{m3}}$, with an additional rotation of the spins towards the basal plane.

### 3.5 $Er_5Pb_3$.

Fig. 16a shows the temperature-dependent magnetic susceptibility $M/H$ (full symbols, left axis) and the inverse susceptibility $H/M$ (open symbols, right axis) for $Er_5Pb_3$ in an applied field $H$ = 0.1 T. The magnetic order is marked by a peak in M/H close to 36 K and, together with the indistinguishable ZFC/FC $M(T)$ close to this temperature, suggests that the transition corresponds to AFM order. The Néel temperature $T_N$ determined from $d(MT)/dT$ is $T_N$ = 36.3 K. Upon further cooling, another local magnetization maximum at $T_2$ = 11.0 K is observed in $d(MT)/dT$, consistent with a spin reorientation transition. The inverse magnetic susceptibility displayed in Fig. 16 (right axis) is linear above 50 K, indicative of Curie-Weiss behavior. From the linear fit of $H/M$ above $T$ = 50 K, the resulting Weiss temperature is $\theta_W$ = 21.3 K, with the effective moment $m_{eff}^{exp} \approx 9.62\ \mu_B/Er^{3+}$. This is consistent with the calculated value $\mu_{eff}^{theory} \approx 9.58\ \mu_B/Er^{3+}$. As was also the case for R = Dy and Ho, the positive $\theta_W$ value indicates FM coupling between $Er^{3+}$ ions, which together with likely strong CEF effects, results in the AFM order. As was the case for the other $R$ members of this series, the ordering temperature $T_{ORD}$ exceeds the Weiss temperature, and in this case $T_{ORD}/\theta_W \sim 2$.

The $T$ = 2 K $M(H)$ isotherm (Fig. 16b) displays a metamagnetic transition with the critical field

$H_c$ = 1.25 (2.37) T for increasing (decreasing) applied field, as determined from the peak position in *dM/dH*. The metamagnetism is likely associated with spin-flop transitions caused by strong CEF anisotropy in this system. It was shown[33] that in antiferromagnets with strong uniaxial anisotropy, *e.g.* hexagonal or tetragonal systems with an easy axis, the transition would be a single-step from a low magnetization state toward a fully saturated state. However, within our field range the maximum magnetization (~ 8 $\mu_B/Er^{3+}$) is less than the $Er^{3+}$ saturated moment $\mu_{sat}$ = 9 $\mu_B/Er^{3+}$. The finite slope of M(H) at $H$ = 7 T suggests that this system is approaching saturation at a field slightly higher than our maximum measured field.

Fig. 17 shows a comparison of the NPD patterns for $Er_5Pb_3$ collected at $T$ = 150, 60, 20 and 4 K, using wavelength λ = 1.54 Å and plotted in the $Q$ = 0.4 - 2.6 $Å^{-1}$ interval. Fig. 18 shows the LeBail refinement of the NPD data for $Er_5Pb_3$ measured at $T$ = 20 K (a) and 4 K (b). The insets show NPD data for $Q$ = 0.1-1.3 $Å^{-1}$. Below $T$ = 20 K the NPD data (Fig. 18a) reveals new magnetic Bragg reflections compared to the high temperature data ($T$ = 150 K, Fig. 17) indicative of AFM state. This is in good agreement with the magnetization data with AFM ordering below $T_N$ = 36.3 K. Indexing the new magnetic peaks suggests that the magnetic unit cell is more than three times larger than the nuclear unit cell with $a_M = a_N$, $c_M = 3.40c_N$. The corresponding propagation magnetic vector determined from the Lebail fit to the $T$ = 20 K NPD data (Fig. 18a) is $\mathbf{k_{m1}}$ = (0, 0, 0.294). A second magnetic transition observed in the magnetization data at $T_2$ = 11 K is caused by an additional spin rotation along the *c axis* (Fig. 18b). At $T$ = 4 K, intensities of the $(h,h,0)+\mathbf{k_m}$ magnetic peaks decrease compared to those measured at $T$ = 20 K, typical of a spin rotation from an in-plane to an out-of-plane arrangement. In our case, out-of-plane spin arrangements lower the magnetic component of the system along *c* axis which leads to the intensity decrease of the $(h,h,0)+\mathbf{k_m}$ magnetic peaks.

### 3.6 $Tm_5Pb_3$.

Fig. 19a shows the temperature-dependent magnetic susceptibility *M/H* (black circles, left axis) and the inverse susceptibility *H*/*M* (open circles, right axis) for $Tm_5Pb_3$ in an applied field of $H$ = 0.1 T. The magnetic order is marked by a peak in M/H close to 18 K and, together with the indistinguishable ZFC/FC $M(T)$ close to this temperature, suggests that the transition corresponds to AFM order. The Néel temperature $T_N$ determined is from $d(MT)/dT$ to be $T_N$ = 17.8 K. The inverse magnetic susceptibility displayed in Fig. 19a (right axis) is linear above 16 K, indicative of Curie-Weiss behavior. From the linear fit of *H*/*M* above $T$ = 40 K, the resulting Weiss temperature is $\theta_W$ = -1.7 K, with the effective moment $m_{eff}^{exp} \approx 7.45\ \mu_B/Tm^{3+}$. This is consistent with the calculated value $\mu_{eff}^{theory} \approx 7.56\ \mu_B/Tm^{3+}$. The negative $\theta_W$ value is consistent with AFM coupling between $Tm^{3+}$ ions.

The $M(H)$ isotherm measured at $T = 2$ K (Fig. 19b) displays two-step metamagnetic transition with a critical field for increasing (decreasing) applied field, as determined from the peaks in the $dM/dH$. 3.5 $\mu_B/Tm^{3+}$ The magnetization in an applied field $H = 7$ T corresponds to 7 $\mu_B/Tm^{3+}$ and is consistent with the calculated saturated moment $\mu_{sat} = 7\ \mu_B/Tm^{3+}$.

A comparison of the NPD patterns for $Tm_5Pb_3$ collected at $T$ = 150, 50, 30 and 4 K (Fig. 20) using wavelength $\lambda = 1.54$ Å plotted in the $Q = 0.70$ - 2.8 Å$^{-1}$ interval reveal extra peaks at $T = 4$ K, suggesting AFM magnetic order between 30 and 4 K. Indexing these magnetic peaks suggests that the magnetic unit cell is more than three times larger than the nuclear unit cell with $a_M = a_N$, $c_M = 3.60c_N$. The corresponding propagation magnetic vector determined from the Lebail fit to the $T = 4$ K NPD data (Fig. 20) is $\mathbf{k_{m1}}$ = (0, 0, 0.275). Such a magnetic vector is similar to the one found in $Er_5Pb_3$ or $Er_5Si_3$.[23]

## IV. CONCLUSIONS

We have successfully synthesized polycrystalline $R_5Pb_3$ ($R$ = Gd-Tm) samples and performed structural and magnetic analysis, along with neutron powder diffraction on the $R$ = Tb – Tm $R_5Pb_3$ compounds. The inverse magnetic susceptibilities of all $R_5Pb_3$ showed that the ordering temperature $T_{ORD}$ decreases when going from $Gd^{3+}$ towards $Tm^{3+}$ (Fig. 21a), consistent with the expected deGennes scaling[ref]. Below $T_{ORD}$, the behavior is complex and reveals other phase transitions at $T_2$, $T_3$, $T_4$. A fit of the linear parts of the H/M to the Curie-Weiss law yields large differences between Weiss temperatures $\theta_W$ and actual measured $T_{ORD}$ (Fig. 21), with the latter several times larger than the former We believe this to be a result of strong CEF effects. Although the Curie-Weiss fit yields positive $\theta_W$ values for $R_5Pb_3$ ($R$ = Gd - Er), NPD analysis indicates more complex ordered states. This is plausibly also caused by different exchange interactions between the two inequivalent $R$ atomic sites, and is often accompanied by the competition between magnetic incommensurability and commensurability.[13-19]

The $T$ = 2 K isotherms for $R_5Pb_3$ with $R$ = Tb - Tm compounds reveals metamagnetic behavior associated with the crystal field anisotropy. For $R_5Pb_3$ (R = Tb and Dy), the $S$-shaped isotherms and their smooth variation in the measured field range is typical of a continuous metamagnetic transition, which is generally of second order. Note that there is an almost nonexistent hysteresis in $R_5Pb_3$ ($R$ = Tb and Dy) isotherms. The opposite was observed for $Er_5Pb_3$. The isotherm for $Er_5Pb_3$ shows a step-like metamagnetic transition with pronounced hysteresis. Such discontinuous behavior is usually, but not always, associated with a first-order transition. The same isotherms for $R_5Pb_3$ with $R$ = Ho and Tm show multiple metamagnetic transitions. Due to the polycrystallinity and air-sensitivity of our samples, one cannot fully describe the interionic interactions between $R$ sites. It is evident that the CEF, frustration (hexagonal structure with two inequivalent crystallographic $R$ sites), and magnetic coupling play a leading role in the complex magnetic properties observed in $R_5Pb_3$.

The ordering temperature Tord follow the expected de Gennes scaling Tord ~ dG, where the deGennes factor dGis given by $dG = (g_J - 1)^2 J(J + 1)$, in which $g_J$ is the Landé factor and $J$ is the total angular momentum of the $R^{3+}$ ion Hund's rule ground state multiplet. The magnetic ordering temperatures ($T_{ORD}$, $T_2$, $T_3$, $T_4$), Weiss temperatures $\theta_W$ and the ratio $f = |\theta_W| / T_{ORD}$ for $R_5Pb_3$ ($R$ = Gd - Tm) are shown in Fig. 21. The dashed line in Fig 21a shows the expected linear de Gennes scaling. CEF effects often reduce the Hund's rule ground state degeneracy, leading to deviations from the Curie-Weiss behavior of the temperature-dependent magnetic susceptibility. In our case, such a behavior manifests itself in $T_{ORD}$ being much higher than the Weiss temperature $\theta_W$. The ratio, $f = |\theta_W| / T_{ORD}$, or the frustration parameter, is usually employed to quantify the frustration in magnetic systems. With frustration, f is close to 10, or at least larger then unity. This ratio in $R_5Pb_3$ ($R$ = Gd – Tm) is notably up to an order of magnitude *smaller* than one, which implies strong crystalline anisotropy in these materials.

Like most of the previously studied $R_5X_3$ compounds ($X$ = Si, Ge, Sn, Sb), complex magnetic structures are also detected in $R_5Pb_3$ ($R$ = Gd - Tm). For the Pb compounds, NPD data reveal the presence of multiple magnetic phases, which appear to be incommensurate for all $R_5Pb_3$. As quoted above, in a large number of the magnetic materials the periodicity near $T_N$ is incommensurate or long period commensurate leading to amplitude modulated magnetic structures. Only a few known compounds exhibit simple commensurate AFM structures over the whole temperature range below $T_N$ (*e.g.* $ErGa_2$).[34] The analysis of our NPD data shows that magnetic propagation vectors $\mathbf{k_m}$ for $R_5Pb_3$ with changes as follows: Tb (1, 1, 1.88), Dy (1, 3.333, 1), Ho (1, 3.46, 1), Er (1, 1, 3.40) and Tm (1, 1, 3.63), indicating that the $R$ = Dy and Ho compounds have magnetic spins aligned in the *ab*-plane, while in the $R$ = Tb, Er, Tm magnetic spins are aligned along the *c* axis. At low temperatures $T$ = 4 K, a spin rotation occurs in the $R$ = Tb, Ho and Er members of this series, which was observed in other rare earth intermetallics.[33] In these systems, except in the case of a non-magnetic CEF singlet ground state, amplitude modulated magnetic structures are proved not to be stable at low temperature, and additional spin rotation or reorientation is observed.[33]

Based on the increasing *c/a* ratio (Table. 2), it is obvious that the *a* axis decreases more compared to the *c* axis with heavier *R* ions. In other words, upon decreasing the *R* ionic size, the in-plane $d_{R2\text{-}R2}$ distance decreases more than the out-of-plane $d_{R2\text{-}R2}$ distance. This possibly decreases the frustration and allows magnetic spins to order within the *ab*-plane in some $R_5Pb_3$. Thus by geometry, increasing in-plane anisotropy should be expected going from Tb towards Tm. But as one can see from neutron data (Table 1) only $Dy_5Pb_3$ and $Ho_5Pb_3$ show in-plane magnetic ordering. In the other $R_5Pb_3$ (*R* = Tb, Er and Tm) the magnetic spins are aligned along the *c* axis. Neutron data and the magnetic **k** vectors reveal that CEF effects play a very important role. For the smaller members of the series, *R* = Er and Tm, correlations along the *c* axis are much stronger then for example in the *R* = Tb compound, which is why the former compounds have larger correlation lengths (more that 3 unit cells along the *c* axis).

As one can see from Tables 1 and 2, there is no obvious trend that could explain the complex magnetic behavior observed in $R_5Pb_3$; rather the *c/a* ratio, the temperature ratio *f* and the **k** vectors have to be considered together to describe the competition of the in-plane and out-of-plane $d_{R2\text{-}R2}$ magnetic interactions. The character of the modulations and detailed analysis of the NPD data together with the representational symmetry analysis will be published separately.[32] Inelastic neutron scattering and single crystals of the $R_5Pb_3$ are needed in order to determine the crystal field parameters, which would describe quantitatively the complex magnetic behavior observed in $R_5Pb_3$ compounds.

## ACKNOWLEDGEMENT

Work at Rice was partially supported by NSF Grant No. DMR 0847681 and the DOD PECASE. Research conducted at ORNL's High Flux Isotope Reactor was sponsored by the Scientific User Facilities Division, Office of Basic Energy Sciences, US Department of Energy.

Table 1: Measured and calculated magnetic moment, ordering temperatures, Weiss temperature, and incommensurate magnetic structures for $R_5Pb_3$ series.

| $R_5Pb_3$ | $\mu_{eff}$ ( $\mu_B$ per $R^{3+}$) | | $T_{ORD}$ (K) | Weiss temperature $\theta_W$ (K) | Magnetic structure |
|---|---|---|---|---|---|
| | measured | theoretical | | | |
| $Gd_5Pb_3$ | 7.46 | 7.94 | $T_N = 275$<br>$T_2 = 85.5$<br>$T_3 = 4.9$ | 158.7 | No NPD data |
| $Tb_5Pb_3$ | 9.66 | 9.72 | $T_N = 215.3$<br>$T_2 = 69.4$<br>$T_3 = 4$ | 115 | $k_{m1} = (0, 0, 0.535)$<br>$k_{m2} = (0, 0, 0.520)$ |
| $Dy_5Pb_3$ | 10.38 | 10.63 | $T_N = 160.6$<br>$T_2 = 75.3$<br>$T_3 = 43.3$<br>$T_4 = 15.4$ | 36.4 | -*<br>-*<br>-*<br>$k_m = (0, 0.301, 0)$ |
| $Ho_5Pb_3$ | 10.61 | 10.95 | $T_? = 111$<br>$T_2 = 23.7$<br>$T_3 = 8.0$<br>$T_4 = 4$ | 30.4 | $k_{m1} = (0, 0.295, 0)$<br>$k_{m2} = (0, 0.288, 0)$<br>$k_{m2} + k_{m3} = (0, 0.5, 0)$<br>$k_{m2} + k_{m3}$ + spin rotation towards *ab*-plane |
| $Er_5Pb_3$ | 9.62 | 9.58 | $T_N = 36.6$<br>$T_2 = 11$ | 21.3 | $k_{m1} = (0, 0, 0.294)$<br>$k_{m1}$ + spin rotation towards along *c*-axis |
| $Tm_5Pb_3$ | 7.45 | 7.56 | $T_N = 17.8$ | -1.7 | $k_{m1} = (0, 0, 0.275)$ |

* see text (3.3)

Table 2: Room temperature lattice constants, c/a ratio, and all *R*-*R* bond distances for $R_5Pb_3$ series.

| $R_5Pb_3$ | Gd | Tb | Dy | Ho | Er | Tm |
|---|---|---|---|---|---|---|
| ***a* (Å)** | 9.0879(4) | 9.0104(7) | 8.9612(5) | 8.9100(6) | 8.8724(2) | 8.8286(3) |
| ***c* (Å)** | 6.6487(5) | 6.5943(5) | 6.5641(4) | 6.5326(5) | 6.5148(2) | 6.49017(2) |
| ***c/a*** | 0.7316 | 0.7319 | 0.7325 | 0.7332 | 0.7343 | 0.7351 |
| $d_{R1\text{-}R1}$ (Å)$^{+}$ | 3.3182(4) | 3.2962(3) | 3.2813(3) | 3.2628(2) | 3.2474(3) | 3.2329(4) |
| $d_{R1\text{-}R2}$ (Å) | 3.8998(4) | 3.9058(2) | 3.8583(2) | 3.8390(2) | 3.8338(2) | 3.7977(5) |
| $d_{R2\text{-}R2}$ (Å), out-of-plane$^{*}$ | 3.9729(4) | 3.9066(2) | 3.8831(2) | 3.8831(2) | 3.8544(2) | 3.8565(3) |
| $d_{R2\text{-}R2}$ (Å), in-plane$^{*}$ | 3.7844(4) | 3.6319(3) | 3.6880(3) | 3.6469(2) | 3.5963(3) | 3.6417(3) |

$^{+}$ $d_{R1\text{-}R1}$ bonds are depicted in Fig. 2c

$^{*}$ $d_{R2\text{-}R2\ out\text{-}of\text{-}plane}$ (dash light blue line) and $d_{R2\text{-}R2\ in\text{-}plane}$ (solid light blue line) bonds are depicted in Fig. 2d

FIGURE CAPTIONS

Fig. 1 Rietveld analysis of powder x-ray diffraction data of $Ho_5Pb_3$ (fit statistics: $wR_P$ = 7.9%, $R_P$ = 6.0% and $\chi^2$ = 1.4). Collected data are shown as black crosses, calculated model as red line, and background as green line. The Bragg markers are for the $Ho_5Pb_3$ main phase, black asterisk for a Pb impurity.

Fig. 2 **(a)** Left axis: crystallographic *a*-axis versus $r_R^{3+}$ for $R_5Pb_3$. Right axis: the *c* axis versus $r_R^{3+}$. **(b)** Left axis: crystallographic *c*/*a*-ratio plotted versus $r_R^{3+}$. Right axis: volume plotted versus $r_R^{3+}$. The solid lines are the guides to the eye. Depicted crystal structure of $R_5Pb_3$ consists of **(c)** linear chains of $R_1$ atoms along *c* axis (pink solid arrow highlights $d_{\mathrm{R1-R1}}$ bond distance); and **(d)** distorted $R_2$ hexagonal rings and $R_1$ honeycomb layers (dark blue line). Also depicted are in-plane $d_{\mathrm{R2-R2}}$ (solid light blue line) and out-of-plane $d_{\mathrm{R2-R2}}$ (dash light blue line) bond distances.

Fig. 3: **(a)** Left axis: temperature dependence of the magnetic susceptibility of $Gd_5Pb_3$ measured in different magnetic fields H = 0.1 T, 1 T, and 7 T. Right axis: inverse magnetic susceptibility. Solid black line is a Curie-Weiss fit. The inset shows a magnetic transition observed at low temperatures. **(b)** dM/dT for $Gd_5Pb_3$ measured at $H$ = 0.1 T. Inset displays a d$MT$/d$T$ for $Gd_5Pb_3$ measured at $H$ = 1 T.

Fig. 4: The field-dependent magnetization measured at T = 2 K for $Gd_5Pb_3$. The inset displays The field-dependent magnetization measured at low magnetic fields.

Fig. 5: **(a)** Left axis: ZFC and FC temperature dependence of the magnetic susceptibility of $Tb_5Pb_3$ measured in magnetic field H = 0.1 T. Right axis: inverse magnetic susceptibility. The solid red line is a Curie-Weiss fit. **(b)** The magnetization isotherm M(H) measured at T = 2 K. The inset shows a hysteresis observed at low magnetic fields ($H$ < 2.2 T).

Fig. 6: Comparison of the NPD patterns for $Tb_5Pb_3$ collected at different temperatures using wavelength of $\lambda$ = 1.54 Å. The inset shows the NPD data measured at $T$ = 2 K using wavelength $\lambda$ = 2.41 Å in the range of $Q$ = 0.2 - 1.8 Å$^{-1}$ revealing the most intense (0,0,1)+$\mathbf{k_{m2}}$ magnetic peak at $Q$ ~ 0.5 Å$^{-1}$.

Fig. 7: **(a)** The temperature evolution of the (2,1,1)+$\mathbf{k_m}$ magnetic Bragg reflection, as a most intense magnetic peak, revealing the ordering temperature $T_{ORD}$ of $Tb_5Pb_3$. An inset depicts the FM contribution on top of the nuclear peaks (black arrows) by overlaying data collected at $T$ = 300 K (black line) $T$ = 150 K (red line).

Fig. 8: LeBail refinement of the NPD data for $Tb_5Pb_3$ measured at **(a)** T = 60 K and **(b)** 4 K. The difference between the measured data (black) and LeBail fit (red) is shown as a blue line, the calculated Bragg positions are indicated by vertical markers (upper - nuclear phase, lower - magnetic phase). The insets show the difference in incommensurability of the magnetic structures with a magnetic propagation vector **(a)** $\mathbf{k_{m1}}$ = (0,0,0.535) at $T$ = 60 K and **(b)** $\mathbf{k_{m2}}$ = (0,0,0.520) at $T$ = 4 K.

Fig. 9: **(a)** Left axis: ZFC and FC temperature dependence of the magnetic susceptibility of $Dy_5Pb_3$ measured in magnetic field H = 0.1 T. Right axis: dependence of the inverse magnetic susceptibility for $Dy_5Pb_3$. The solid red line is a Curie-Weiss fit. **(b)** The magnetization isotherm M(H) measured at T = 2 K. The inset shows d$M$/d$H$.

Fig. 10: **(a)** Comparison of the NPD patterns collected at different temperatures for $Dy_5Pb_3$. Black arrows indicate magnetic Bragg peaks. The inset shows the temperature evolution of the two magnetic Bragg peaks indicated by black arrows. **(b)** The temperature evolution of the (-3,1,1)+$\mathbf{k_m}$ Bragg reflection, as a most intense magnetic Bragg peak. The solid black line as a polynomial fit and serves as a guide to the eye.

Fig. 11: LeBail refinement of the NPD data for $Dy_5Pb_3$ measured at T = 4 K. The difference between the measured data (black) and LeBail fit (red) is shown as a blue line, the calculated Bragg positions are indicated by vertical markers (upper - nuclear phase, lower - magnetic phase). The inset shows the leBail fit of the two magnetic Bragg peaks.

Fig. 12: **(a)** Left axis: ZFC and FC temperature dependence of the magnetic susceptibility of the $Ho_5Pb_3$ measured in magnetic field H = 0.1 T. Right axis: dependence of the inverse magnetic susceptibility for $Ho_5Pb_3$. The solid red line is a Curie-Weiss fit. **(b)** The magnetization isotherm M(H) measured at T = 2 K.

Fig. 13: **(a)** The temperature evolution of the (3,-1,0)+$\mathbf{k_{m1}}$ Bragg reflection, as a most intense magnetic Bragg peak revealing the ordering temperature $T_{ORD}$ of $Ho_5Pb_3$. **(b)** The temperature evolution of the (1,0,0)+$\mathbf{k_{m2}}$ Bragg reflection revealing the second magnetic transition $T_2$ of $Ho_5Pb_3$. An inset shows the temperature evolution of the (0,1,0)+$\mathbf{k_{m2}}$+$\mathbf{k_{m3}}$ magnetic Bragg reflection revealing the third magnetic transition $T_3$ of $Ho_5Pb_3$.

Fig. 14: LeBail refinement of the NPD data for $Ho_5Pb_3$ measured at $T$ = 50 K using magnetic propagation vector $\mathbf{k_m}$ = (0,0.296,0). The difference between the measured data (black) and LeBail fit

(red) is shown as a blue line, the calculated Bragg positions are indicated by vertical markers (upper - nuclear phase, lower - magnetic phase). The inset shows a comparison of the NPD patterns for $Ho_5Pb_3$ collected at temperatures $T$ = 150 K (purple) and 300 K (black) plotted in $Q$ = 0-0.75 Å$^{-1}$ and $Q$ = 1.6-2.8 Å$^{-1}$ interval.

Fig. 15: Temperature evolution of the NPD data for $Ho_5Pb_3$ collected at temperatures **(a)** $T$ = 50 K (brown), 15 K (orange) and **(b)** $T$ = 15 K (orange), and 4 K (blue) plotted in **(a)** $Q$ = 0.6-1.9 Å$^{-1}$ and **(b)** $Q$ = 1.5-3.5 Å$^{-1}$ interval.

Fig. 16: **(a)** Left axis: ZFC and FC temperature dependence of the magnetic susceptibility of the $Er_5Pb_3$ measured in magnetic field H = 0.1 T. Right axis: dependence of the inverse magnetic susceptibility for $Er_5Pb_3$. The solid red line is a Curie-Weiss fit. **(b)** magnetization isotherm M(H) measured at T = 2 K.

Fig. 17: Comparison of the NPD patterns collected at different temperatures for $Er_5Pb_3$.

Fig. 18: **(a)** LeBail refinement of the NPD data for $Er_5Pb_3$ measured at $T$ = 20 K and **(b)** at $T$ = 4 K using magnetic propagation vector $k_m$ = (0,0,0.294). The difference between the measured data (black) and LeBail fit (red) is shown as a blue line, the calculated Bragg positions are indicated by vertical markers (upper - nuclear phase, lower - magnetic phase). The insets show NPD data for $Er_5Pb_3$ plotted in an interval $Q$ = 0.1-1.3 Å$^{-1}$ at **(a)** $T$ = 20 K and **(b)** $T$ = 4 K. Arrows highlight the intensity difference of the low $Q$ ($h$,$h$,0) magnetic peaks (see text).

Fig. 19: **(a)** Left axis: ZFC and FC temperature dependence of the magnetic susceptibility of the $Tm_5Pb_3$ measured in magnetic field H = 0.1 T. Right axis: temperature dependence of the inverse magnetic susceptibility for $Tm_5Pb_3$. The solid red line is a Curie-Weiss fit. **(b)** The magnetization isotherm M(H) measured at T = 2 K.

Fig. 20: LeBail refinement of the NPD data for $Tm_5Pb_3$ measured at T = 4 K using magnetic propagation vector $k_m$ = (0,0,0.275). The difference between the measured data (black) and LeBail fit (red) is shown as a blue line, the calculated Bragg positions are indicated by vertical markers (upper - nuclear phase, lower - magnetic phase). The inset shows the comparison of the NPD patterns collected at different temperatures for $Er_5Pb_3$ plotted in $Q$ = 0.7-2.6 Å$^{-1}$ interval.

Fig. 21: **(a)** Magnetic ordering temperatures (left axis) and Weiss temperatures $\theta_W$ (right axis). The dashed line is a guide to the eye and shows the expected de Gennes scaling. **(b)** Frustration parameter $f = |\theta_W| / T_{ORD}$.

Fig. 1.

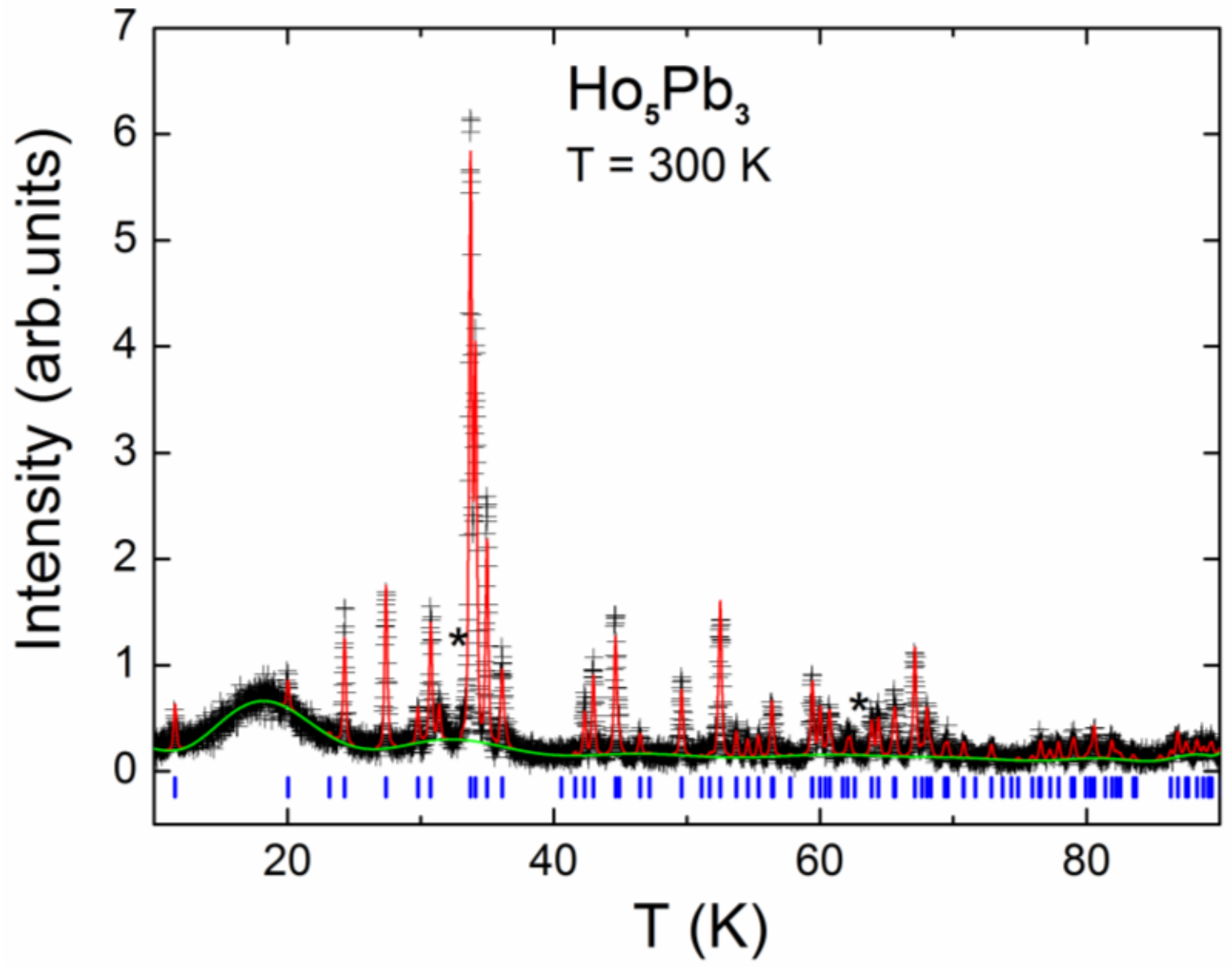

Ho5Pb3
T = 300 K
Intensity (arb.units)
T (K)

Fig. 2.

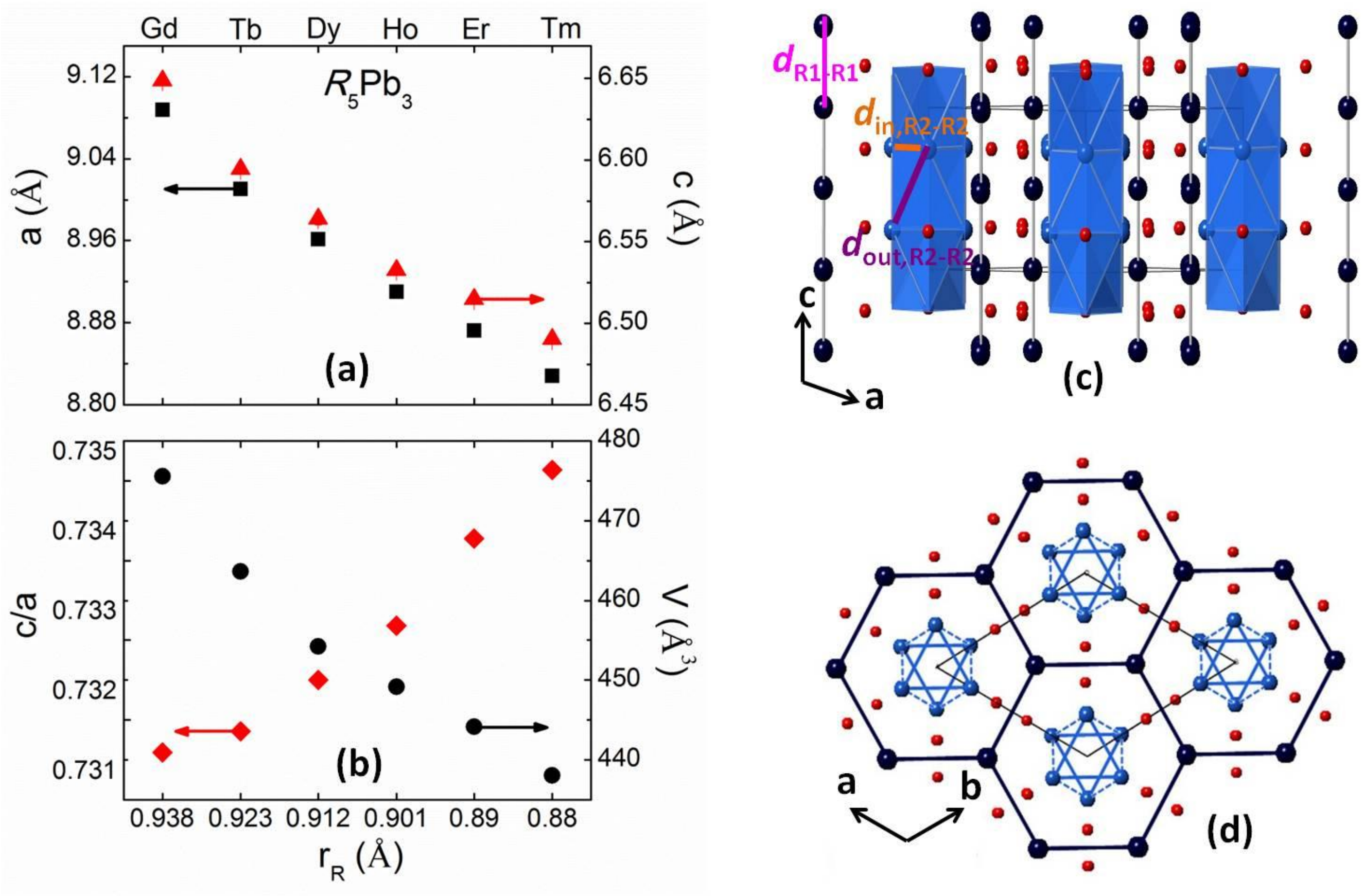

FIG. 3.

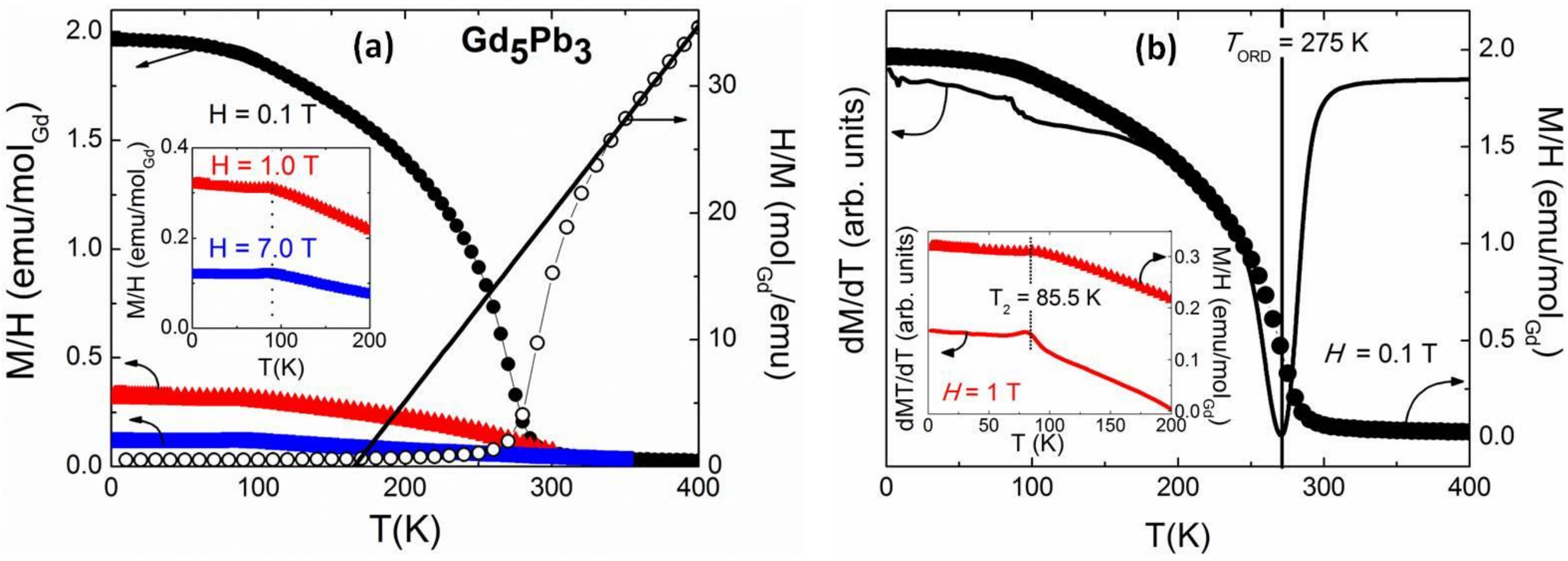

(a)
Gd5Pb3
H = 0.1 T
H = 1.0 T
H = 7.0 T
M/H (emu/molGd)
H/M (molGd/emu)
T(K)
(b)
TORD = 275 K
dM/dT (arb. units)
M/H (emu/molGd)
T2 = 85.5 K
dMT/dT (arb. units)
H = 1 T
H = 0.1 T
T (K)

Fig.4

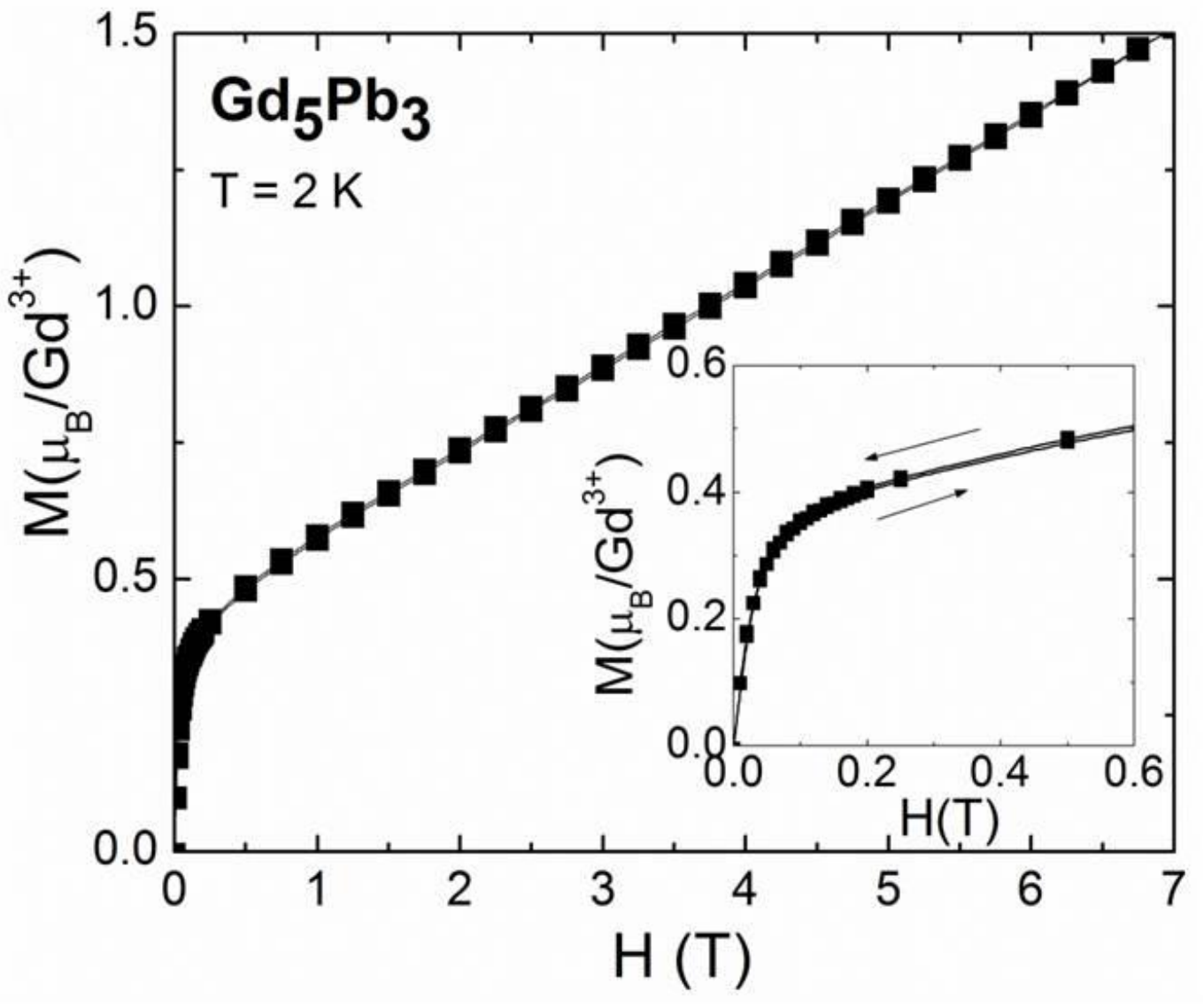

Gd5Pb3
T = 2 K
M(μB/Gd3+)
H (T)
H(T)

FIG. 5.

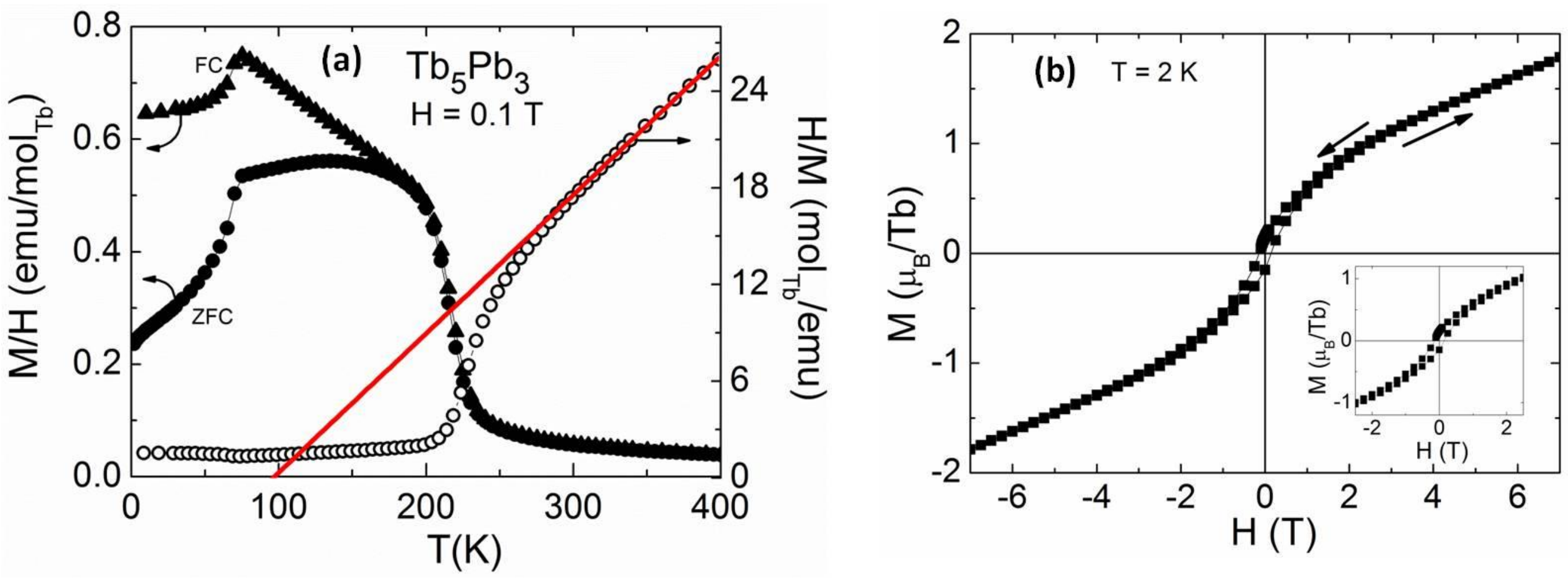

FIG. 6

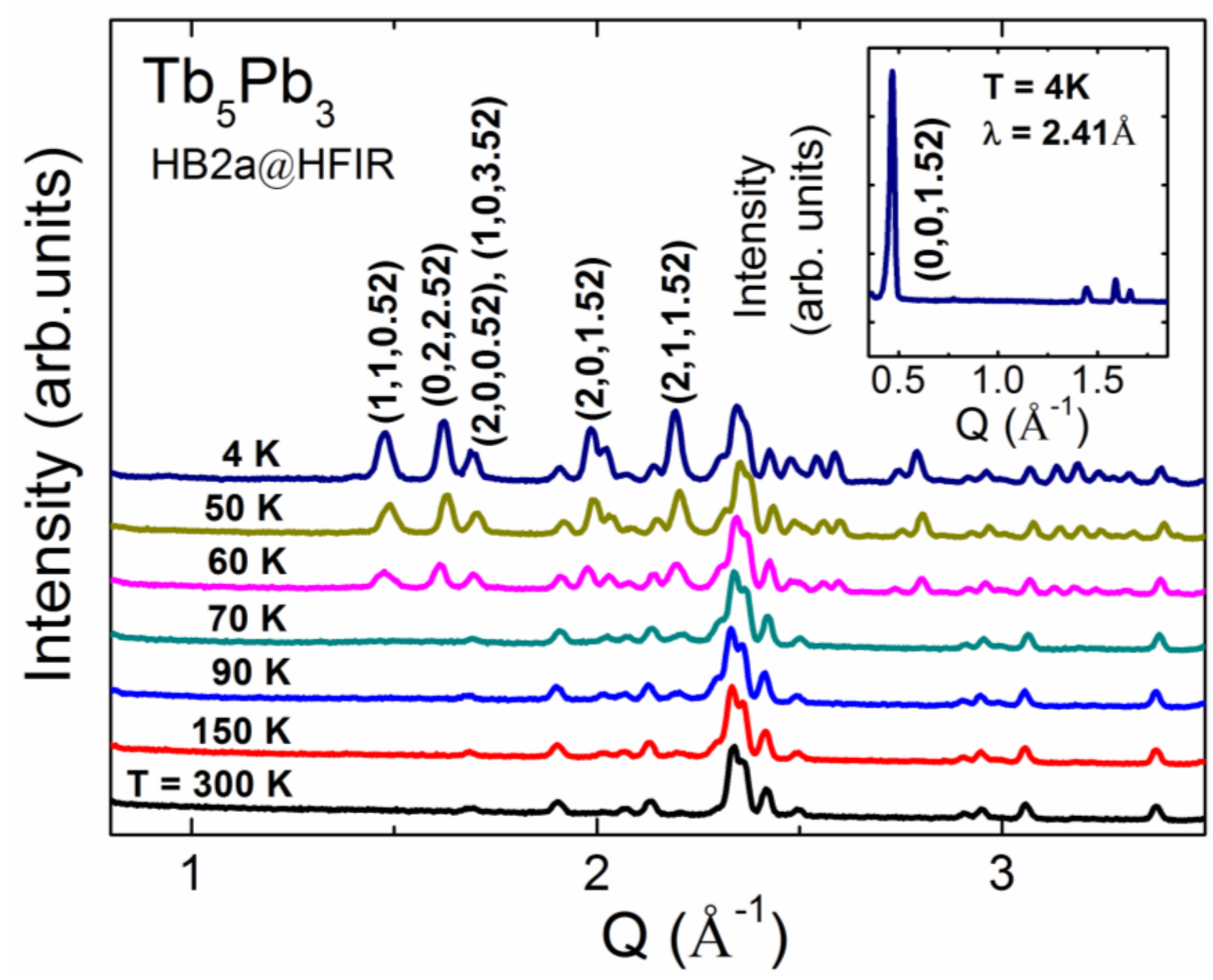

FIG. 7

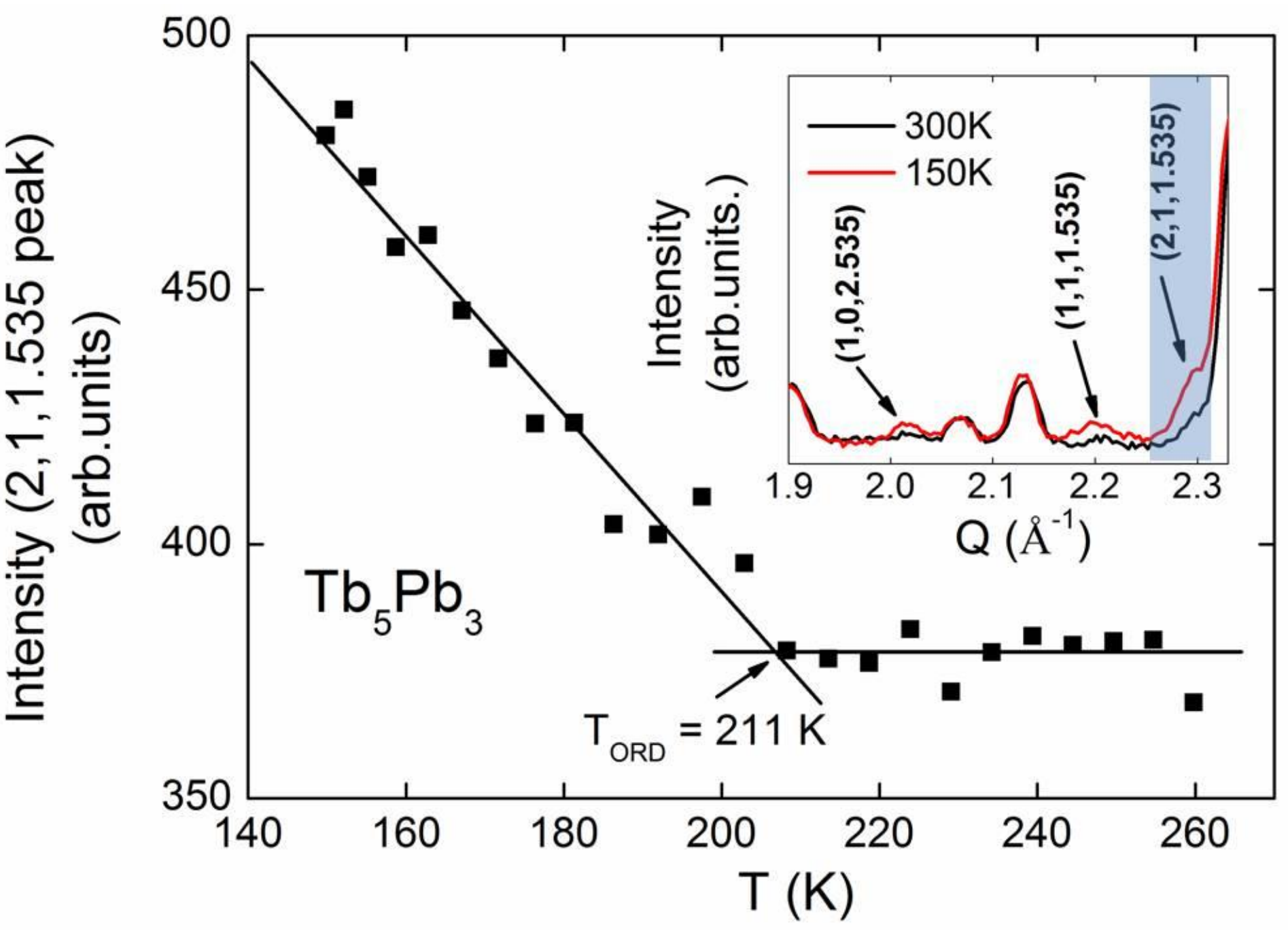

FIG. 8

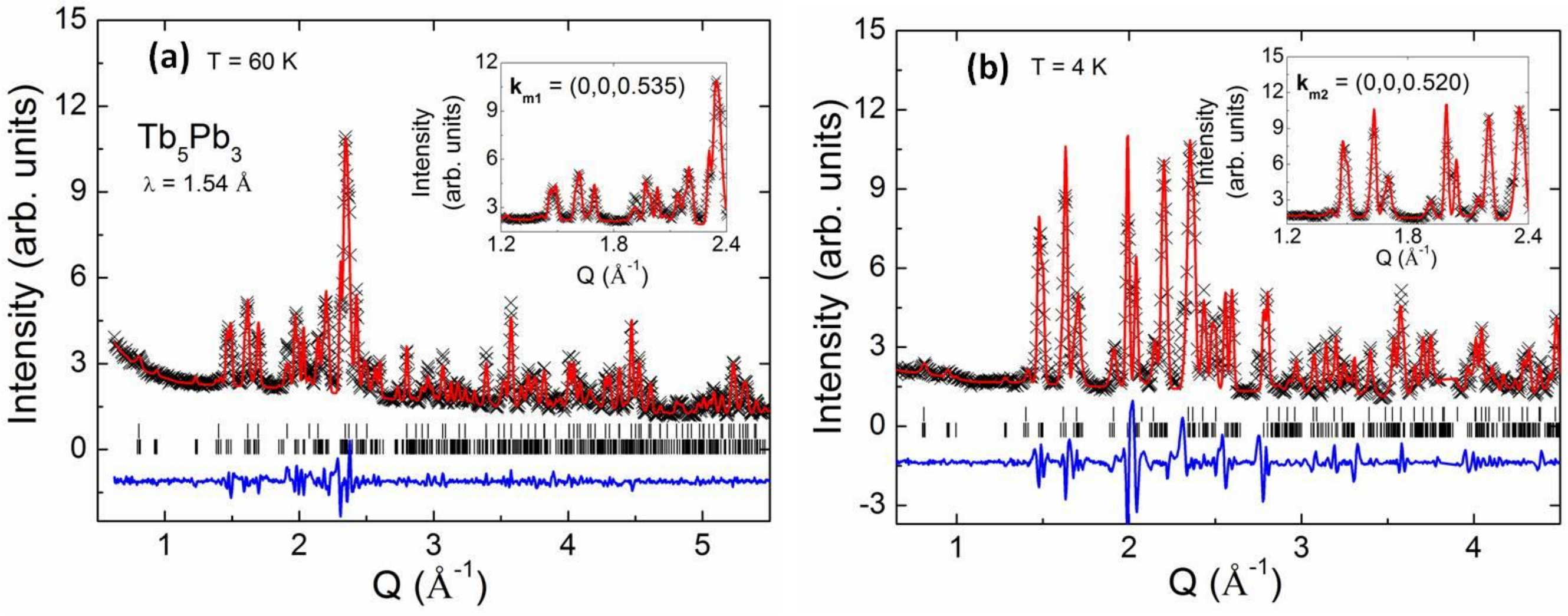

(a) T = 60 K
Tb5Pb3
λ = 1.54 Å
km1 = (0,0,0.535)
(b) T = 4 K
km2 = (0,0,0.520)
Intensity (arb. units)
Q (Å-1)

FIG. 9

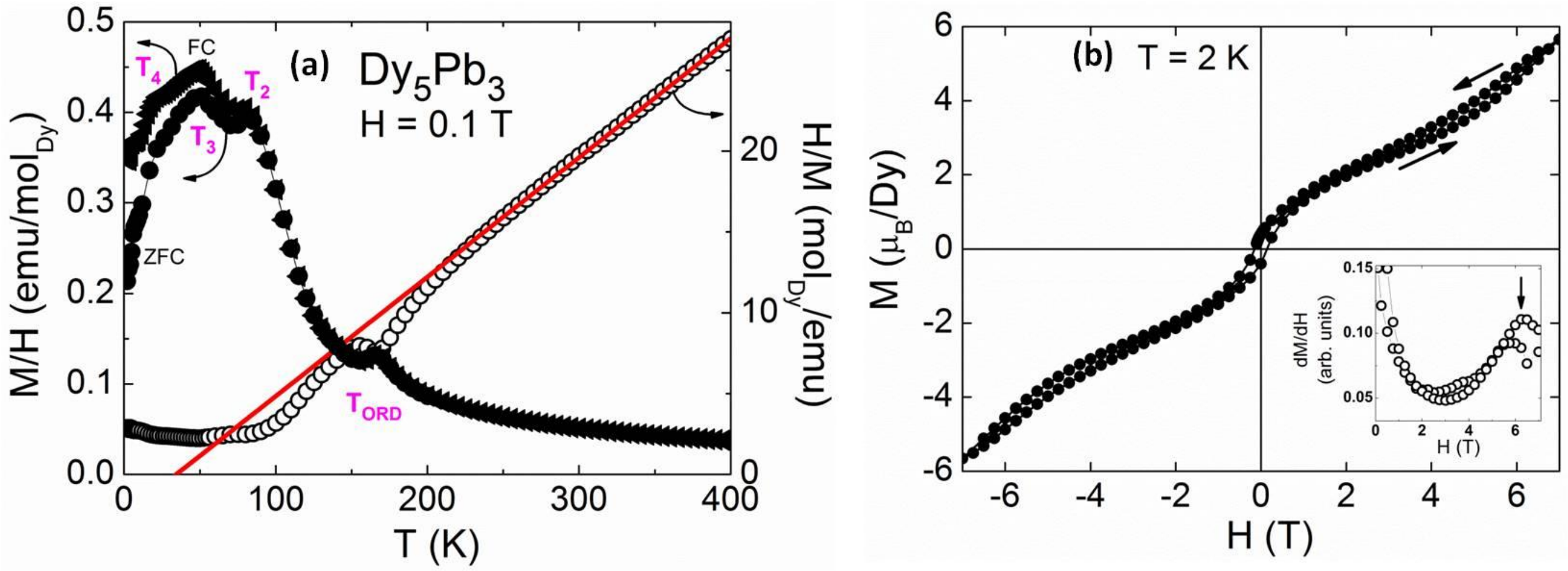

FIG. 10

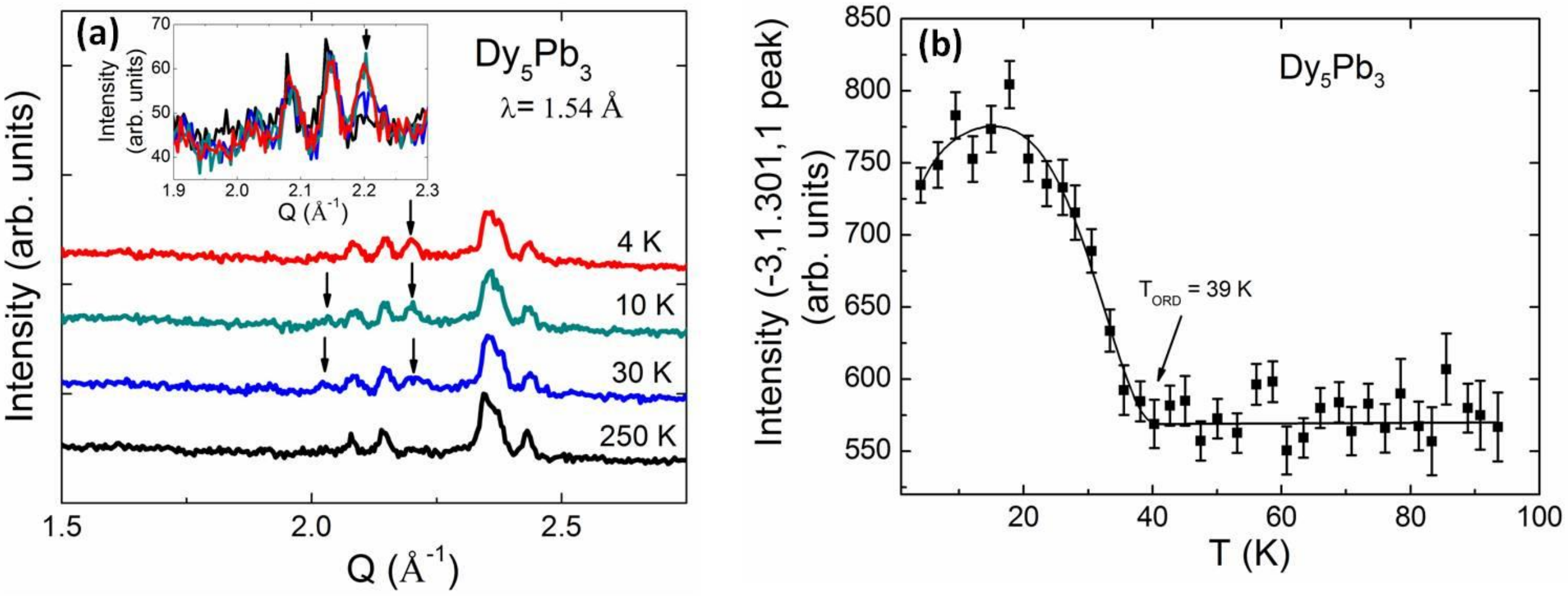

FIG. 11

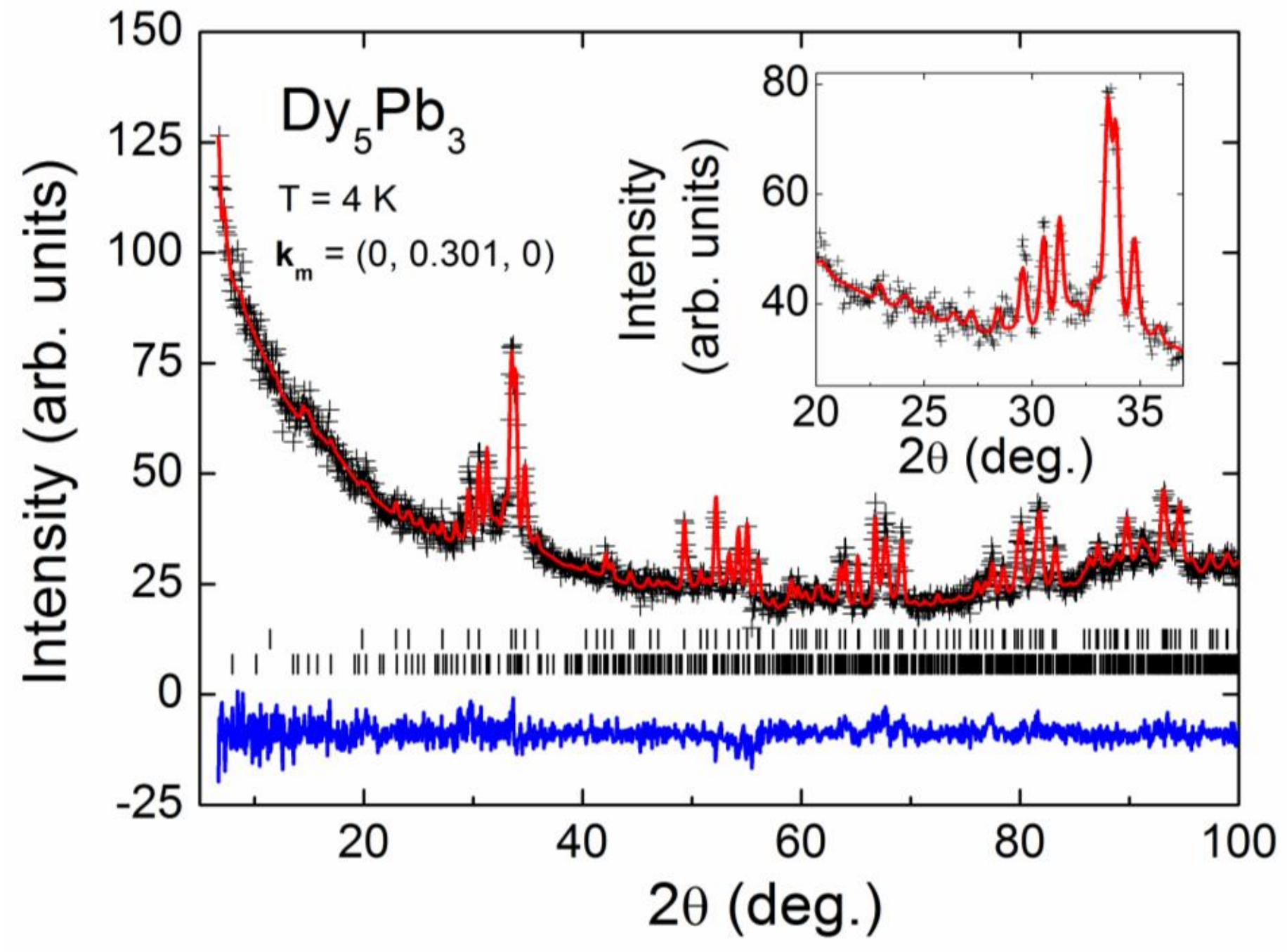
Dy5Pb3
T = 4 K
km = (0, 0.301, 0)
Intensity (arb. units)
2θ (deg.)

FIG. 12

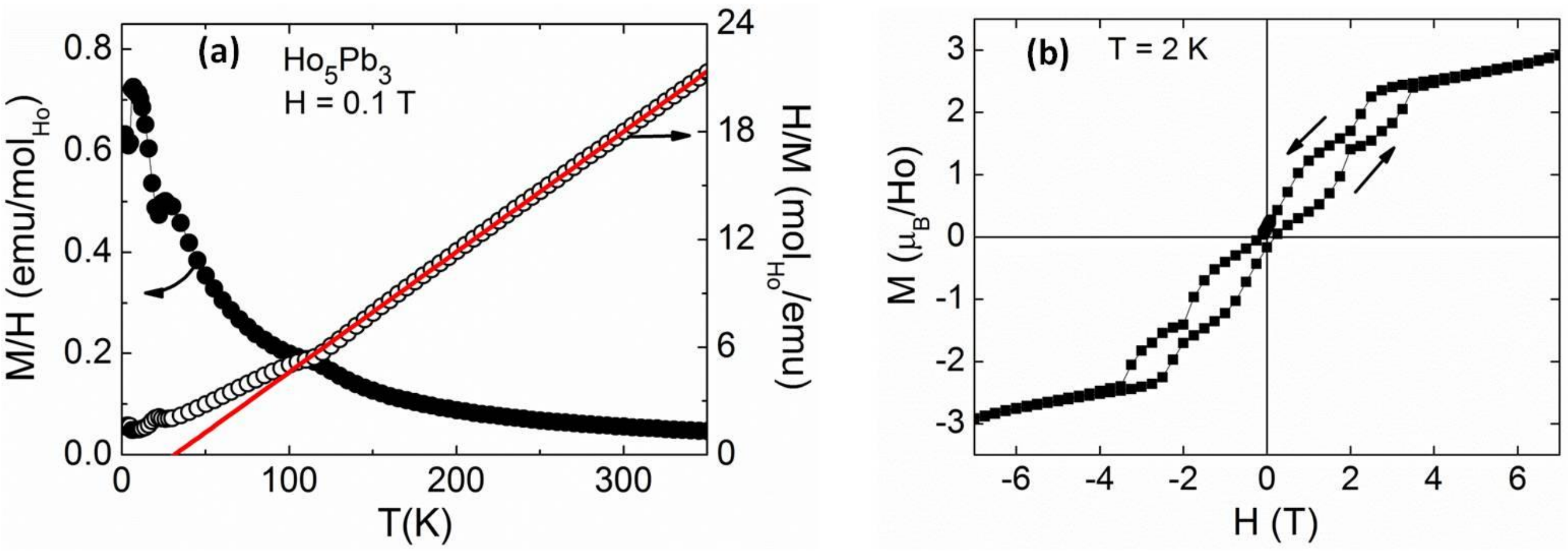

FIG. 13

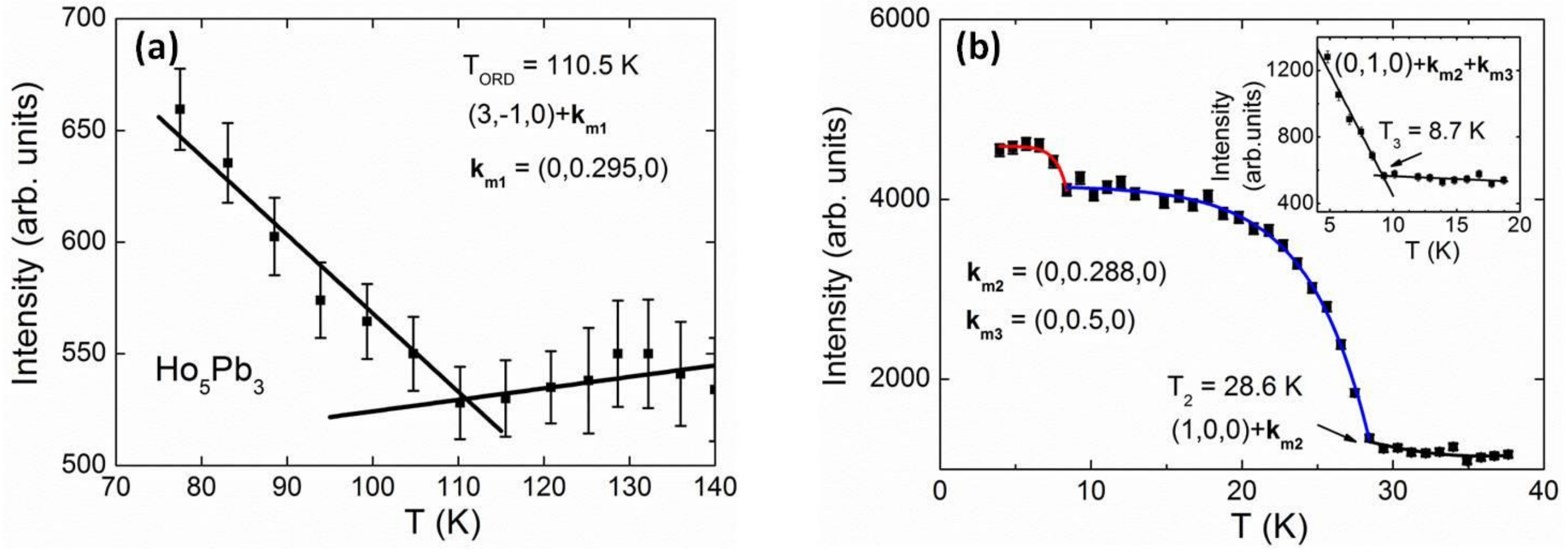

FIG. 14

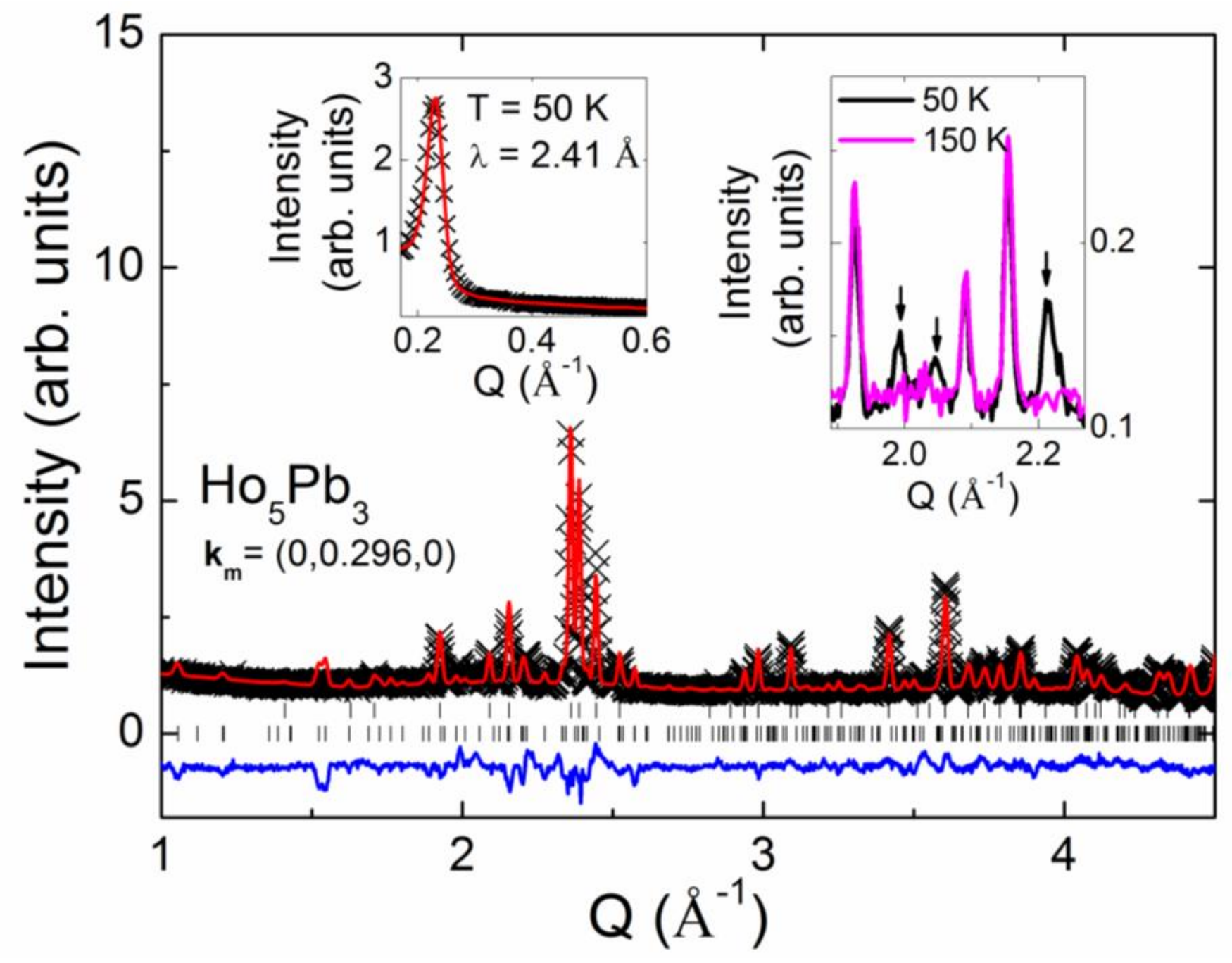

FIG. 15

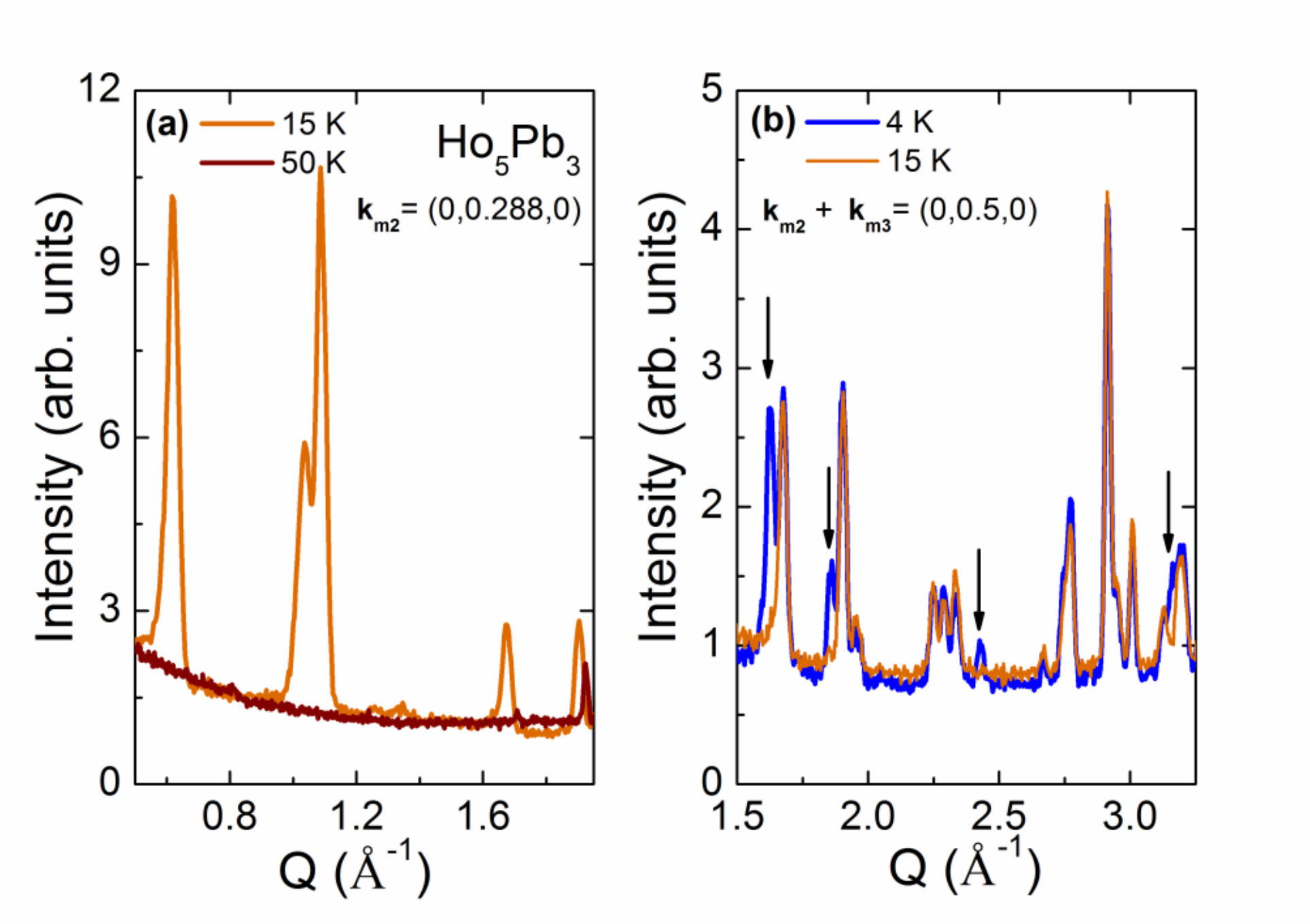

FIG. 16

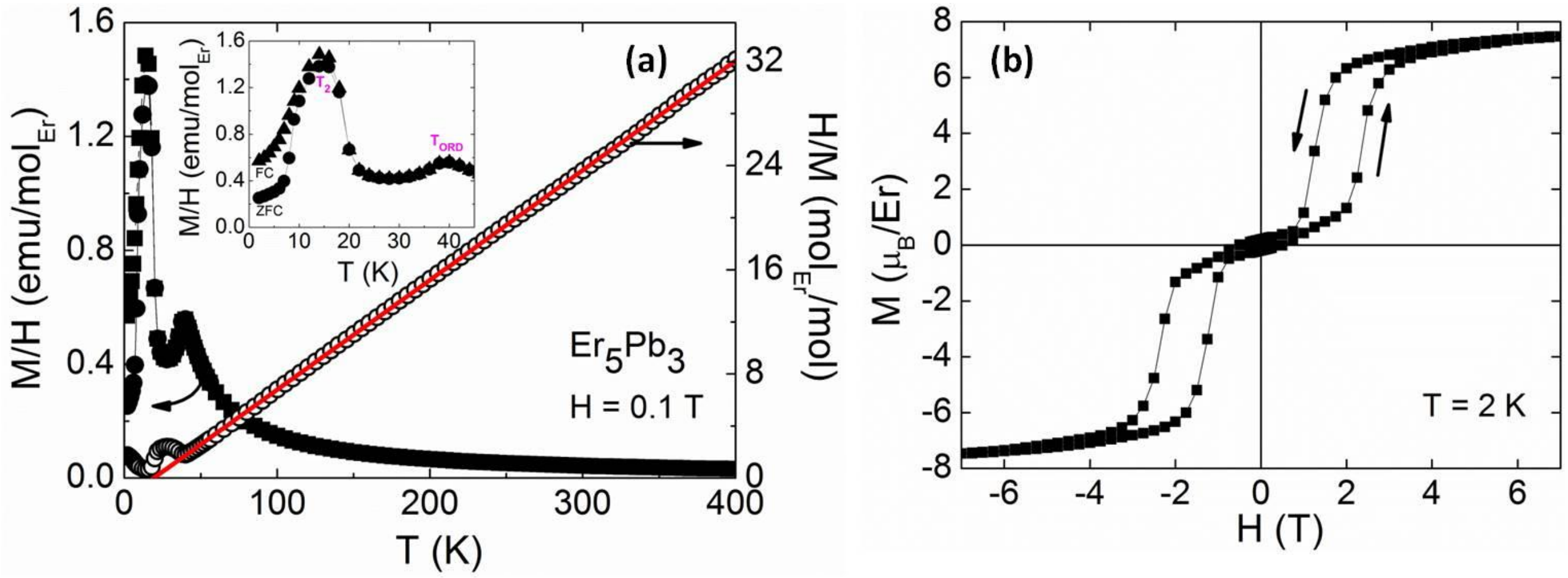

FIG. 17

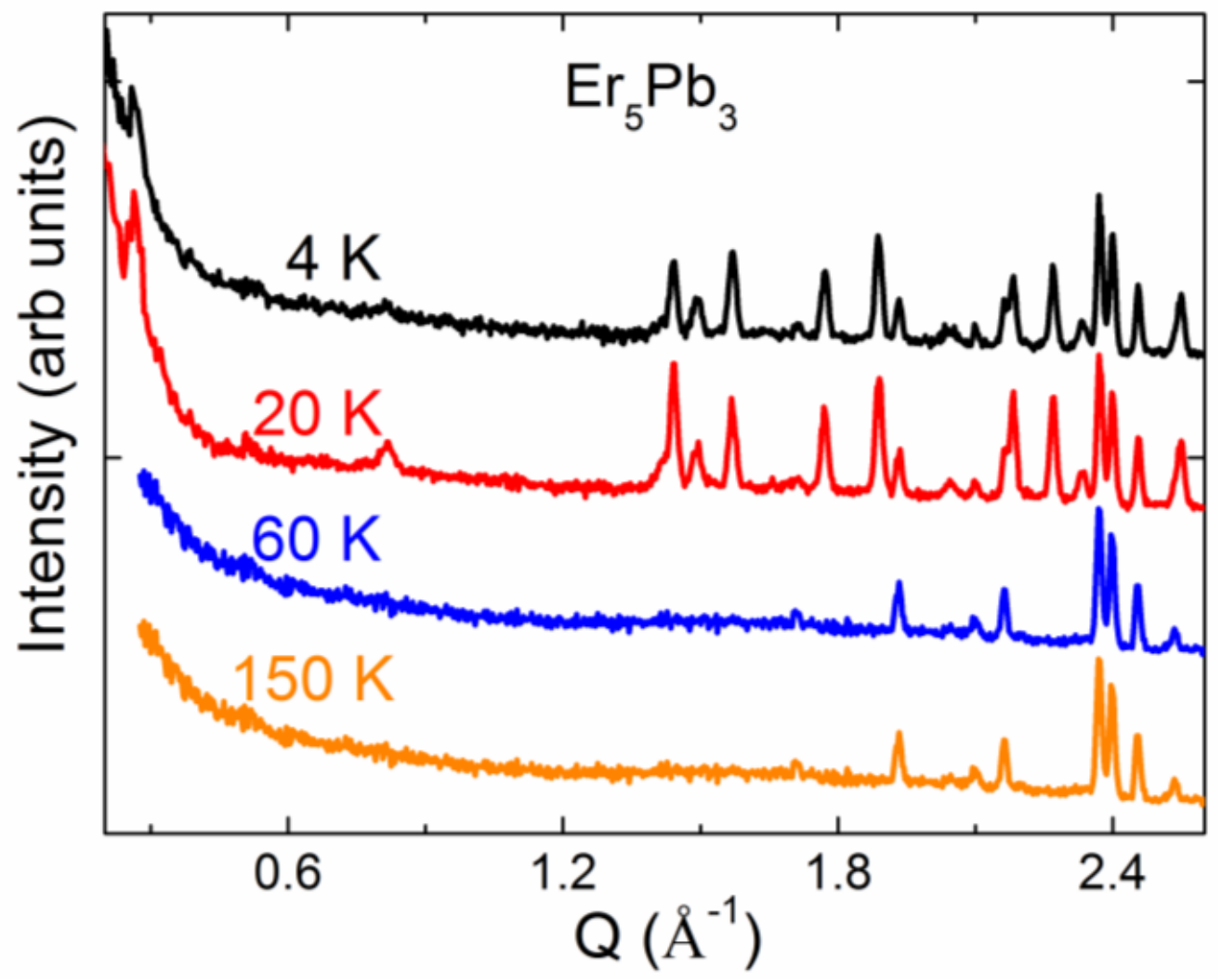

FIG. 18

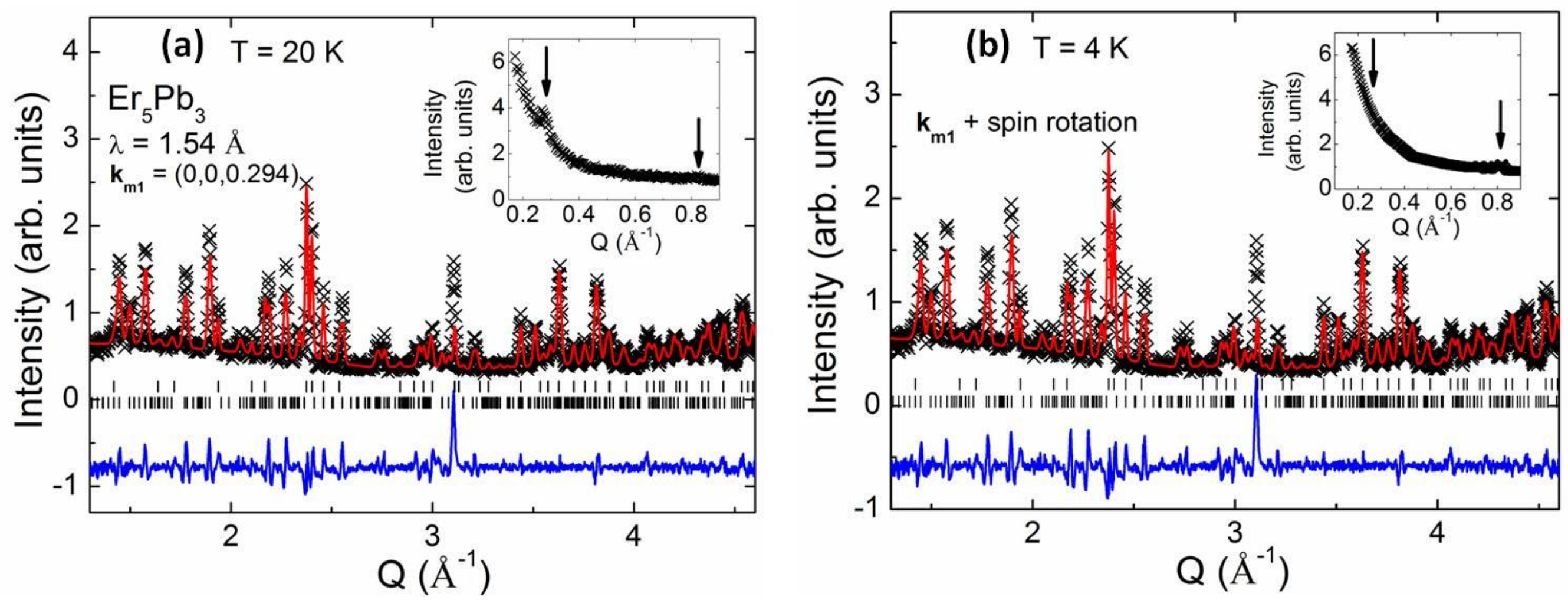

(a)
T = 20 K
$Er_5Pb_3$
λ = 1.54 Å
$\mathbf{k}_{m1}$ = (0,0,0.294)
Intensity (arb. units)
Q (Å$^{-1}$)
(b)
T = 4 K
$\mathbf{k}_{m1}$ + spin rotation
Intensity (arb. units)
Q (Å$^{-1}$)

FIG. 19

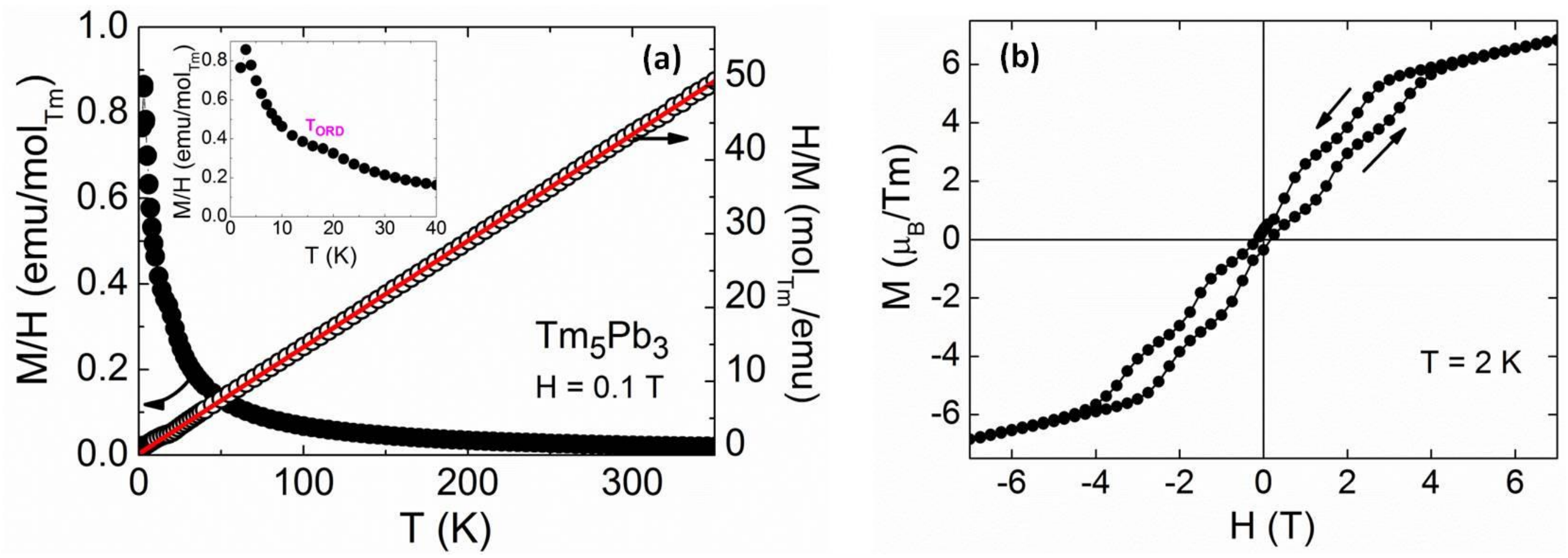

FIG. 20

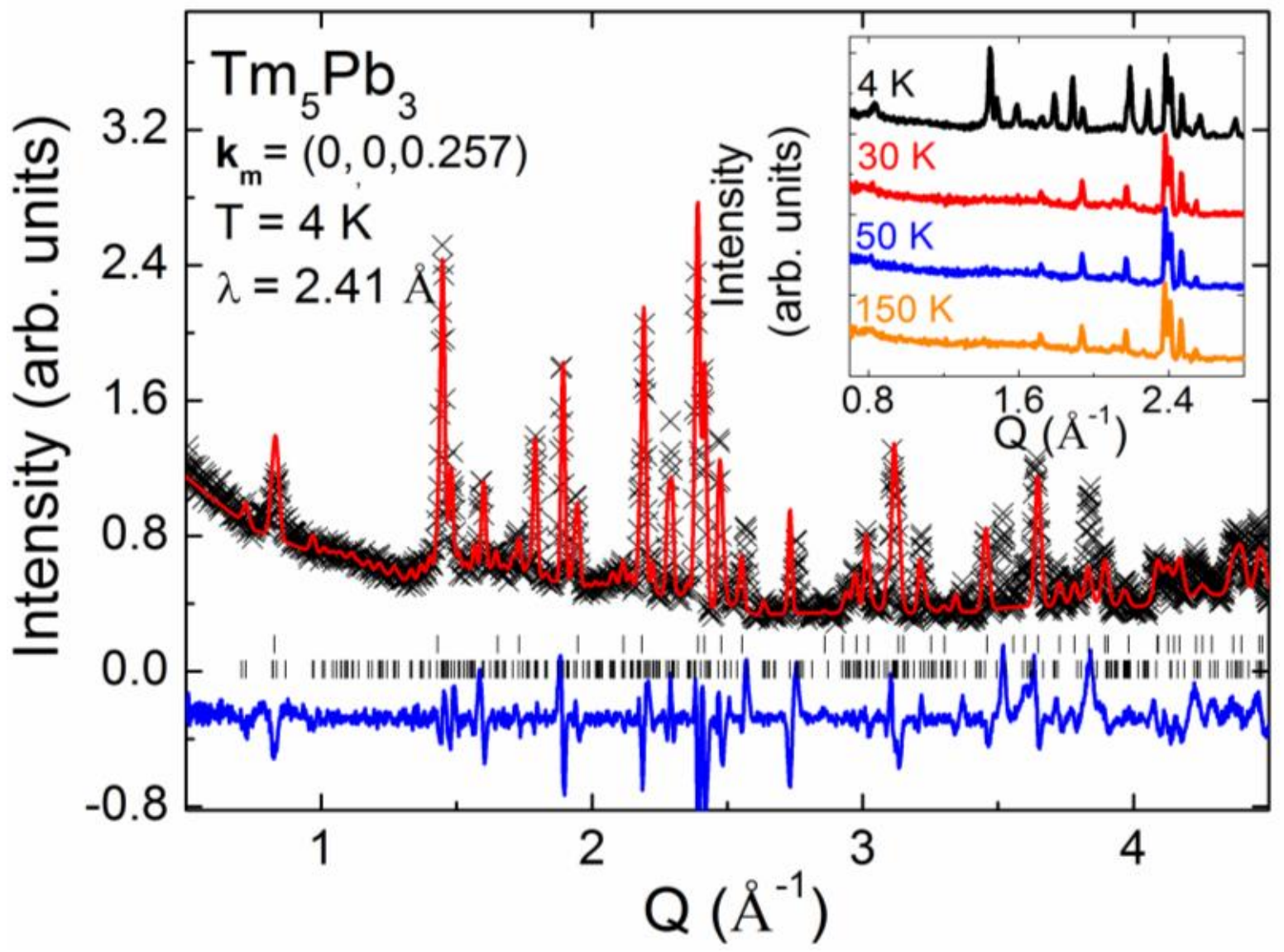

FIG. 21

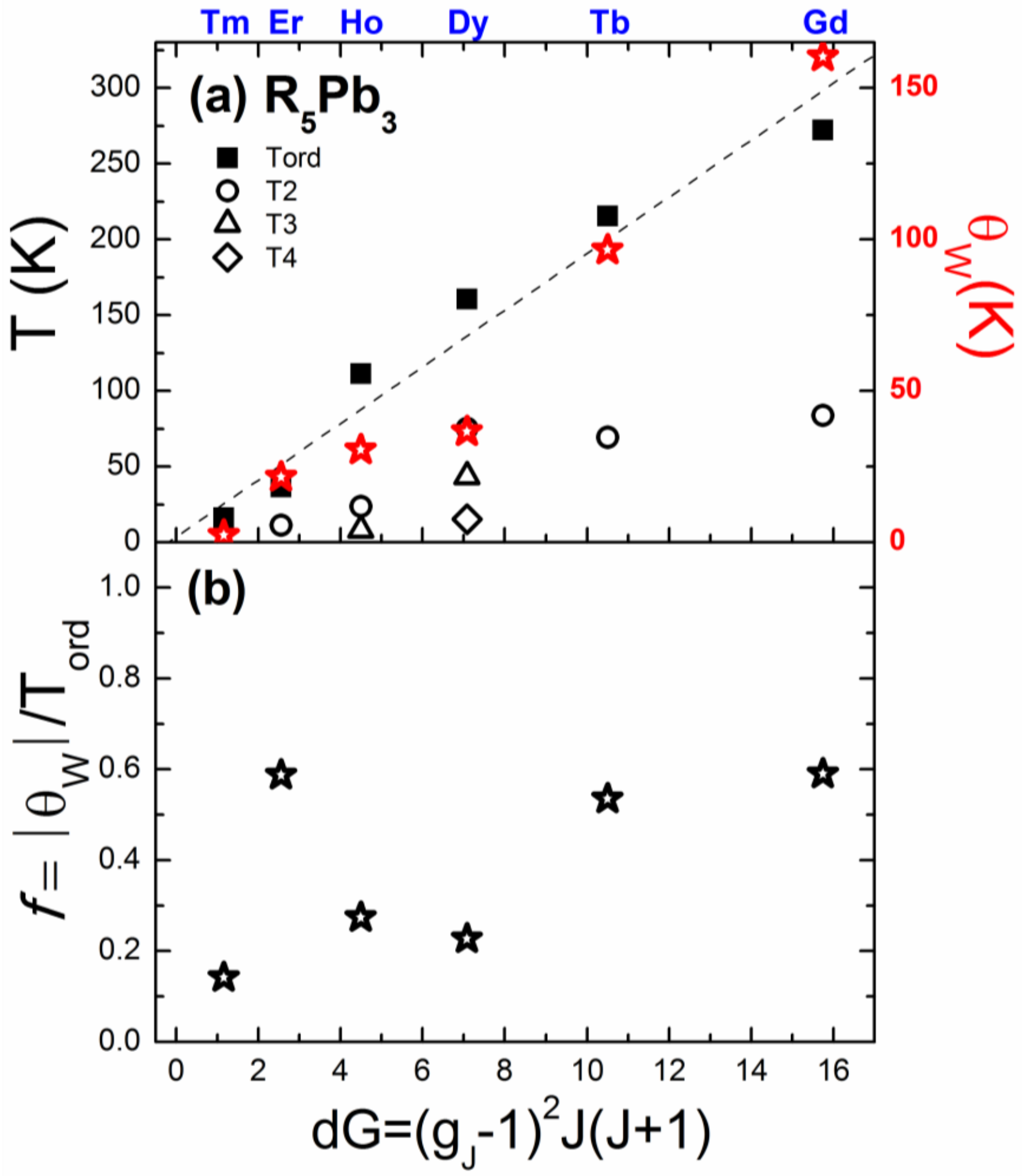